\documentclass[preprint,twocolumn]{aastex62}
\usepackage{amsmath}
\usepackage{mathtools}
\usepackage{natbib}
\usepackage{threeparttable}
\usepackage{booktabs}
\usepackage{bibentry}

\def\eg{{\it e.g.,}}
\def\ie{{\rm i.e.,}}

\def\logr{ {log$_{10}$(R/R$_*$)} }
\newcommand{\be}{\begin{equation}}
\newcommand{\ee}{\end{equation}}

\newcommand{\flash}{{\sc FLASH}}
\newcommand{\maihem}{{\sc MAIHEM}}
\newcommand{\cloudy}{{\sc Cloudy}}

\newcommand{\HI}{\ion{H}{1}}
\newcommand{\HII}{\ion{H}{2}}
\newcommand{\HEI}{\ion{He}{1}}

\newcommand{\HEIII}{\ion{He}{3}}
\newcommand{\CI}{\ion{C}{1}}
\newcommand{\CII}{\ion{C}{2}}
\newcommand{\CIII}{\ion{C}{3}}
\newcommand{\CIV}{\ion{C}{4}}
\newcommand{\CV}{\ion{C}{5}}

\newcommand{\CVII}{\ion{C}{7}}
\newcommand{\NI}{\ion{N}{1}}

\newcommand{\NV}{\ion{N}{5}}

\newcommand{\NVIII}{\ion{N}{8}}
\newcommand{\OI}{\ion{O}{1}}

\newcommand{\OVI}{\ion{O}{6}}

\newcommand{\OIX}{\ion{O}{9}}
\newcommand{\NEI}{\ion{Ne}{1}}

\newcommand{\NEVII}{\ion{Ne}{7}}
\newcommand{\NEVIII}{\ion{Ne}{8}}

\newcommand{\NEXI}{\ion{Ne}{11}}
\newcommand{\NAI}{\ion{Na}{1}}

\newcommand{\NAVI}{\ion{Na}{6}}
\newcommand{\MGI}{\ion{Mg}{1}}
\newcommand{\MGII}{\ion{Mg}{2}}

\newcommand{\MGVI}{\ion{Mg}{6}}
\newcommand{\SiI}{\ion{Si}{1}}

\newcommand{\SiIV}{\ion{Si}{4}}

\newcommand{\SiVI}{\ion{Si}{6}}
\newcommand{\SI}{\ion{S}{1}}

\newcommand{\SVI}{\ion{S}{6}}
\newcommand{\ARI}{\ion{Ar}{1}}

\newcommand{\ARVI}{\ion{Ar}{6}}
\newcommand{\CAI}{\ion{Ca}{1}}

\newcommand{\CAVI}{\ion{Ca}{6}}
\newcommand{\FEI}{\ion{Fe}{1}}

\newcommand{\FEVI}{\ion{Fe}{6}}
\newcommand{\ELEC}{${e}^{-}$}

\begin{document}
\title{Non-equilibrium Ionization States Within Galactic Outflows: Explaining Their \OVI\ and \NV\ Column Densities}
\author{William J. Gray}
\affil{CLASP, College of Engineering, University of Michigan, 2455 Hayward St., Ann Arbor, Michigan 48109, USA}
\author{Evan Scannapieco}
\affil{School of Earth and Space Exploration, Arizona State University}
\author{Matthew D. Lehnert}
\affil{Sorbonne Universit\'e, CNRS UMR 7095, Institut d'Astrophysique de Paris, 98 bis bd Arago, 75014 Paris, France}

\begin{abstract}

We present a suite of one-dimensional spherically-symmetric hydrodynamic simulations that study the atomic ionization structure of galactic outflows.
We track the ionization state of the outflowing gas with a non-equilibrium atomic chemistry network that includes photoionization, photo-heating, and ion-by-ion cooling.  
Each simulation describes a steady-state outflow that is defined by its mass and energy input rates, sonic radius, metallicity, and UV flux from both the host galaxy and meta-galactic background.   We find that for a large range of parameter choices, the ionization state of the material departs strongly from what it would be in photo-ionization equilibrium, in conflict with what is commonly assumed in the analysis of observations.  
In addition, nearly all the  models reproduce the low \NV\ to \OVI\ column density ratios and the relatively high \OVI\ column densities that are observed.

\end{abstract}
\keywords{galaxies: evolution -- galaxies: starburst -- methods: numerical -- hydrodynamics -- ISM: jets and outflows}

\section{Introduction}

Galactic outflows are ubiquitous in intensely star-forming galaxies across all redshifts \citep{Heckman1990,Lehnert96,Heckman2000,Pettini2002,Martin2005,Rupke2005,Veilleux2005,Weiner2009,Bordoloi2014,Bordoloi2016,Rubin2014,Heckman2015,Chisholm2016}. 
These outflows are powered by stars through supernovae \citep[\eg][]{Maclow1999,Scannapieco2001,Mori2002,Scannapieco2002,Springel2003,Dalla2008,Creasey2013}, stellar winds \cite[\eg][]{Murray2011,Hopkins2012,Muratov2015,Hayward2017}, and cosmic rays \citep[\eg][]{Farber2018}. 
They have dramatic effects on their host galaxies by decreasing metallicities \citep[\eg][]{Tremonti2004,Oppenheimer2009,Dave2011,Lu2015,Agertz2015},  suppressing star-formation \cite[\eg][]{Somerville1999,Cole2000,Scannapieco2001,Scannapieco2002,Benson2003} and perhaps occasionally enhancing it \cite[\eg][]{Scannapieco2004b,Gray2010,Gray2011a,Gray2011b,Bieri2016,Fragile2017,Mukherjee2018}.

The nature of galactic outflows is very complex, and a complete picture is possible only when they are observed and modeled  over the full range of temperatures -- from X-ray observations of 10$^{7}$ to 10$^{8}$ K material \citep{Martin1999,Strickland2007,Strickland2009}, near-UV and optical observations of warm ionized gas at $\approx$10$^{4}$ \citep[\eg][]{Pettini2001,Tremonti2007,Martin2012,Soto2012}, and IR and submm observations of molecular gas at 10-10$^{3}$ K \cite[\eg][]{Walter2002,Sturm2011,Bolatto2013}.
Most studies of galactic outflows focus on warm gas at $\approx$10$^{4}$K due to the strong emission and absorption lines in the rest-frame UV and optical. 
X-ray observations, however, can only be obtained for nearby objects, since only they are bright enough to be seen \citep[\eg][]{Lehnert99,Strickland2009}. 
This hot gas is fundamentally important as it dominates the thermal and kinetic energy of the outflow and provides the best evidence that winds actually escape their host galaxies \citep[\eg][]{Lehnert99}. 

Since galactic outflows are generally diffuse in nature and thus have low surface brightnesses in their extended emission, spatially-resolved emission line studies are difficult in anything other than nearby galaxies \citep[\eg][]{Lehnert96,Westmoquette2009,Sharp2010,Arribas2014,Spence2018}. 
Therefore, absorption lines are the most suitable choice for studying the kinematics, column density, and mass and momentum outflow rates of winds. 

Unfortunately, several issues complicate any analysis of the properties of winds. 
One of the persistent mysteries is the exact relationship between the hot and cold phases. 
One possibility is that cold material is being driven out of the host galaxy by ram pressure acceleration from hotter material entrainment \citep[\eg][]{Lehnert96, Veilleux2005}. 
However, this hypothesis runs into serious difficulties, both because: (i) shocks and conduction from the exterior medium tend to compress the cloud perpendicular to the direction of the flow, greatly reducing the momentum flux it receives; and (ii) instabilities and evaporation lead to rapid cloud disruption \citep{Klein94, MacLow1994, Orlando2006, Orlando2008, Scannapieco2015, Bruggen2016}.  
A second possibility is that cold gas is formed from the cooling of high temperature, $10^7-10^8$K gas, which is moving at high radial velocities. 
Such scenarios for explaining the cold gas in outflows was first proposed by  \cite{Wang1995a,Wang1995b} and extended to include cooling \citep{Efstathiou2000},  infall and metallicity evolution \citep{Silich2003,Silich2004,Tenorio-Tagle2007,Wunsch2011}, and the formation of density inhomogeneities \citep{Scannapieco2017}. 
This possibility has received significant attention with recent observations of ultra-luminous infrared galaxies (ULIRGs) and other starbursts suggesting that it is plausible \citep{Zhang2015, Martin2015,Thompson2016,Zhang2017}.

Determining outflow properties without this full picture can lead to incorrect estimates in the mass outflow rate \citep{Chisholm2018}. 
For example, accurate rates can only be determined if we understand the ionization state of the outflowing gas. 
Photoionization models \citep[\eg\ \cloudy;][]{Ferland2013} have been used interpret observations of UV absorption lines to estimate outflow rates. 
However, these models assume a steady state in which all the rates of ionizing and recombination are in equilibrium. 

To better understand the impact of non-equilibrium effects, we present here a suite of one-dimensional spherically symmetric simulations that study the hydrodynamic and ionization structure of galactic outflows. 
We implement an inflow boundary condition in the simulation box that reproduces outflow properties of the freely expanding adiabatic outflow conditions at the sonic radius in the model of \cite{Chevalier1985}. 
The ionization state of the gas is computed using a non-equilibrium atomic chemistry package that includes photoionization, photo-heating, and ion-by-ion recombination and cooling. 

The structure of the paper is as follows: \S2 introduces the galactic outflow model. 
\S3 discusses the model framework and initial conditions. 
The results of our simulations are presented in \S4. 
In \S5, we discuss the \NV-\OVI\ ratio and its importance to observations. 
Finally, we summarize and conclude our study in \S6. 

\section{Galactic Wind}

We are interested in the ionization structure of an expanding outflow generated from a starbursting galaxy. 
Since the outflow moves over several to many gas scale heights per Myr and most starbursts last for many Myr \citep[\eg][]{Greggio1998, Forester2003, McQuinn2010}, the outflow is often described as being in equilibrium. Such an assumption seems to match observations when reliable X-ray analyses can be made \citep[\eg][]{Heckman1990,Heckman1995,Ott2005,Strickland2007,Yukita2012}.

For a spherically symmetric outflow and assuming that gravitational forces are negligible the equations of motion become:

\begin{equation}
\label{eqn:mass}
\frac{1}{r^2}\frac{d}{dr}(\rho u r^2) = q_{m},
\end{equation}
\begin{equation}
\label{eqn:mom}
\rho u \frac{du}{dr} = -\frac{dP}{dr}-q_{m}u,
\end{equation}
\begin{equation}
\label{eqn:energy}
\frac{1}{r^2}\frac{d}{dr}\left[\rho u r^2\left(\frac{1}{2}u^2+\frac{\gamma}{\gamma-1}\frac{P}{\rho} \right) \right]=q_{e},
\end{equation}

\noindent
where $r$ is the radial coordinate, $\rho$ is the mass density, $u$ is the radial velocity, $\gamma$ is the adiabatic index, and $P$ is the pressure. 

The mass and energy input rates are:
\begin{equation}
q_{m} = 
\begin{cases}
\dot{M}/V  & \text{if}\ r \le R_{*} \\
0          & \text{if}\ r > R_{*},  \\
\end{cases}
\end{equation}
\begin{equation}
\label{eqn:qe}
q_{e} = \\
\begin{cases}
\dot{E}/V                 & \text{if}\ r \le R_{*} \\
-\Lambda n^2 - \mathcal{H} n   & \text{if}\ r > R_{*},
\end{cases}
\end{equation}
\noindent
where $R_*$ is the sonic radius (see below), $V=4\pi R_{*}^3/3$, $n_i$ is the number density of species $i$, $n_e$ is the electron number density, $\Lambda$ is total cooling function, and $\mathcal{H}$ is the total photo-heating rate.

In the case where $q_e$ and $q_m$ are zero outside the sonic radius the solution to these equations is presented in \cite[][CC85 hereafter]{Chevalier1985}. 
These three parameters, the energy inflow rate, $\dot{E}$, the mass inflow rate, $\dot{M}$, and the sonic radius, $R_{*}$, define the hydrodynamics of the outflow. 
In the adiabatic case presented in CC85, the density, pressure, and velocity asymptotically approach $\rho \sim r^{-2}$, $P\sim r^{-10/3}$, and $u\sim u_{\infty}=\sqrt{2\dot{E}/\dot{M}}$. 
It is with this solution in mind that we perform the simulations presented here and implemented as the inflow boundary conditions in the simulation. 
In the following sections we describe our simulations in detail. 
In particular, the inflow boundary conditions meant to represent the outflow at the sonic radius.

\section{Model Framework and Initial Conditions}

All of the simulations presented here are run with \maihem\ \citep{Gray2015,Gray2016,Gray2017}, our modified version of the adaptive mesh hydrodynamics code, \flash \citep[version 4.5][]{Fryxell2000}.
The atomic package has recently been updated to fully resolve some elements, \ie\ carbon, nitrogen, oxygen, and neon, as well as now including argon.
Our atomics package tracks the non-equilibrium evolution of 84 species across 13 elements: hydrogen (\HI-\HII), helium (\HEI-\HEIII), carbon (\CI-\CVII), nitrogen (\NI-\NVIII), oxygen (\OI-\OIX), neon (\NEI-\NEXI), sodium (\NAI-\NAVI), magnesium (\MGI-\MGVI), silicon (\SiI-\SiVI), sulfur (\SI-\SVI), argon (\ARI-\ARVI), calcium (\CAI-\CAVI), iron (\FEI-\FEVI), and electrons (\ELEC). 
Our atomic network consists of 310 reactions. 
For each species we consider electron impact ionization, radiative and dielectronic recombination, and photoionization due to a UV background. 

To compute the photo-heating and photoionization rates, we utilize photoionization cross sections from \cite{Verner1995} for the inner shell electrons and \cite{Verner1996} for the outer shell electrons. 
The photoionization and photo-heating rates due to the UV background are parameterized by the line intensity at the Lyman limit, $\mathcal{J_{\nu}}=$10$^{-21}$J$_{21}$ erg s$^{-1}$ cm$^{-2}$ Hz$^{-1}$ Sr$^{-1}$, where J$_{21}$ is a free parameter. 

Two improvements have been made to our network since last presented in \cite{Gray2017}. 
First, J$_{21}$ is now defined cell-by-cell rather than as a global variable. 
This provides the ability to define a spatially varying, although temporally fixed, radiation field. 
Second, the photoionization and photoheating rates are computed at the beginning of every simulation. 
The radiation spectrum is defined at runtime via an input file. 
This allows for the flexibility of changing the input spectrum without recompiling the code.

The inclusion of the chemistry network and its constituent species necessitates updates to some of the equations presented above.
The total cooling becomes
\begin{equation}
\Lambda n^2=\sum_i \Lambda_i n_i n_e,
\end{equation}
where $\Lambda_i$ and $n_i$ are the cooling function and number density of species $i$, and $n_e$ is the number of electrons. The cooling rates are computed using \cloudy\ using the same procedure as described in \cite{Gray2015} and \cite{Gray2016}, which has been updated to include the new ions in our expanded network.
Our procedure creates a table of cooling rates that is then linearly interpolated over to compute the cooling rate for a given temperature.
Similarly, the total photoionization rate, $\mathcal{H}$, is
\begin{equation}
\mathcal{H} n = \sum_i \mathcal{H}_i n_i,
\end{equation}
where $\mathcal{H}_i$ is the photoheating rate for ion $i$.

The mass inflow rate, $\dot{M}$, energy inflow rate, $\dot{E}$, and the sonic radius, $R_{*}$ define the properties of the galactic outflow and are used as the basis of our inflow boundary conditions. 
With these conditions, the energy flux of the outflow is then:
\begin{equation}
\dot{E} = \dot{M}\left(1/2 v_{\rm outflow}^2 +3/2 c_{s}^2 \right),
\end{equation}

\noindent
where $v_{\rm outflow}$ is the outflow velocity and c$_s$ is the sound speed. 
At the sonic radius, R$_*$, 

\begin{equation}
c_s^2=v_{\rm outflow}^2=\dot{E}/2\dot{M}.
\end{equation}
The mass density of the outflow is,
\begin{equation}
\Omega R_*^2 \rho v_{\rm outflow} = \dot{M},
\end{equation}

\begin{figure}
\begin{center}
\includegraphics[trim=0.0mm 0.0mm 0.0mm 0.0mm, clip, width=0.95\columnwidth]{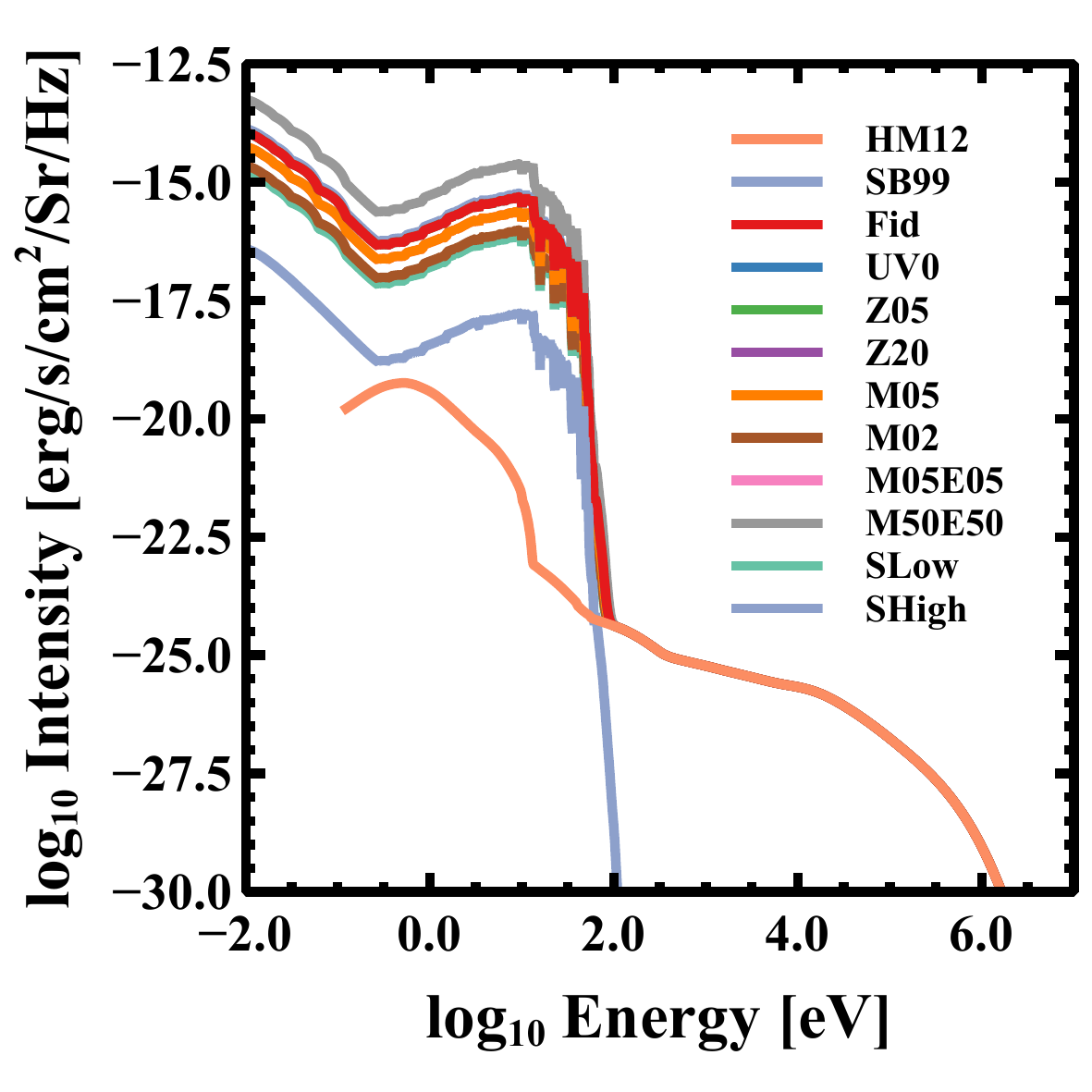}
\caption{ Incident spectra used in our models. The legend gives the line color for the spectra used for the given model. HM12 refers to the standard metal-galactic background of \cite{Haardt2012} while SB99 represents the starburst spectra generated from \cite{Leitherer1999}. All other lines show the composite incident spectra used in our models.}
\label{fig:Spectrum}
\end{center}
\end{figure}

\noindent
where $\Omega R_*^2$ is the effective area of the outflow. 
The density of the outflow is then given by, $\rho = \dot{M}/\Omega R_*^2 v_{\rm outflow}$, 
for completely isotropic outflow, \ie\ $\Omega=$4$\pi$. 
Here we take $\Omega=\pi$, as used in \cite{Scannapieco2017} based on observations of a large sample of low-redshift starbursts \citep{Heckman1990, Lehnert96, Martin2012}. 

The mass input rate can be scaled by the star-formation rate as, $\dot{M}=\beta \dot{M}_{SFR}$, where $\beta$ is the mass-loading factor of the outflow. 
Likewise, the energy input rate is scaled as $\dot{E}=\alpha \dot{E}_{SN}$, where $\alpha$ is the energy-loading factor, which accounts for the fraction of the energy from supernova directed into the outflow.  
If one assumes that a single supernova event generates 10$^{51}$ ergs of energy, and that there is one supernova per 100 M$_\odot$ formed, the total energy input rate is then $\dot{E}=$10$^{49}\alpha \dot{M}_{SFR}$ erg yr$^{-1}$. 

\begin{table*}
\begin{centering}
\resizebox{0.95\textwidth}{!}{%
\begin{threeparttable}
	\label{tab:runsummary}
	\begin{tabular}{|l|cccccccccc|}
	\hline
	Name &  $\dot{M}$        & $\dot{E}$         & $\alpha$ & $\beta$ & V$_{outflow}$ & n           & T            & Z             & UV & SFR           \\
	     &  [M$_{\odot}$/yr] & [10$^{50}$erg/yr] &          &         & [km/s]        & [cm$^{-3}$] & [10$^{6}$ K] & [Z$_{\odot}$] &    & [M$_{\odot}$] \\
	\hline
	\hline
	Nominal  & 10  & 1.0 & 1.0 & 1.0 & 500  & 4.6 & 18 & 1.0 & \checkmark & 10  \\
	UV0      & 10  & 1.0 & 1.0 & 1.0 & 500  & 4.6 & 18 & 1.0 &            & 10  \\
	Z05      & 10  & 1.0 & 1.0 & 1.0 & 500  & 4.6 & 18 & 0.5 & \checkmark & 10  \\
	Z20      & 10  & 1.0 & 1.0 & 1.0 & 500  & 4.6 & 18 & 2.0 & \checkmark & 10  \\
	M05      & 5   & 1.0 & 1.0 & 0.5 & 707  & 1.7 & 36 & 1.0 & \checkmark & 10  \\
	M02      & 2   & 1.0 & 1.0 & 0.2 & 1100 & 0.4 & 91 & 1.0 & \checkmark & 10  \\
	M05E05   & 5   & 0.5 & 0.5 & 0.5 & 500  & 2.3 & 18 & 1.0 & \checkmark & 10  \\
	M50E50   & 50  & 5.0 & 0.5 & 0.5 & 500  & 23  & 18 & 1.0 & \checkmark & 100 \\
	\hline
	SLow     & 1.5 & 0.47& 1.0 & 0.3 & 885  & 0.4 & 57 & 1.0 & \checkmark & 5   \\
	SHigh    & 12  & 1.7 & 0.9 & 0.6 & 594  & 4.7 & 26 & 1.0 & \checkmark & 20  \\

\hline
\end{tabular}
\begin{tablenotes}
\item	{\bf Notes:} Summary of the simulations presented. The first column gives the name for each model. The second and third columns give the mass inflow rate and energy inflow rate respectively. Columns 4 and 5 gives the energy loading and mass loading constants. Columns 6-8 present the resulting outflow velocity, number density, and temperature of the outflow. The ninth column gives the metallicity of the outflow. A checkmark in column 10 shows a model with the UV background implemented. Finally, column 11 gives the star formation rate which is used in determining the magnitude of the UV background, see Eqn.~\ref{eqn:Loutflow}.
\end{tablenotes}
\end{threeparttable}
}
\end{centering}
\end{table*}

The above outflow values are used to implement the boundary condition defined at the inner boundary.
To complete the definition of all hydrodynamic variables, the temperature of the outflow is defined as,
\begin{equation}
\label{eqn:temp}
T_{\rm outflow} \approx v_{\rm outflow}^2m_{H}\bar{A}/ k_{B},
\end{equation}

\noindent
where m$_{H}$ is the mass of hydrogen, $\bar{A}$ is the average atomic weight, and k$_{B}$ is Boltzmann's constant. 
The atomic composition of the gas inflowing at the boundary is assumed to have a solar composition. 
The initial ionization state of the gas is assumed to be equal to the collisional ionization equilibrium (CIE) values that depend only on the outflow temperature and the UV background radiation field. 
The inital CIE values depend on the initial hydrogen number density, input spectra, and ionization parameter.
These values are model dependent and require the generation of a model specific CIE table. 
We employ \cloudy\ to compute these CIE values.
This procedure creates a table of values that which is read in by \flash\ at runtime and interpolated over to give the initial ionization state of the gas at the boundary. 

\subsection{Background UV Radiation Field}

The incident UV background radiation field is a combination of two sources, the meta-galactic UV background from \cite{Haardt2012} and a starburst model \citep[SB99;][]{Leitherer1999}. 
The SB99 models are run with fixed stellar mass, with a metallicity of one-tenth solar, and the spectra that are evolved for $t=$4$\times$10$^{4}$ years. 
We have compared spectral energy distributions at earlier times and with solar metallicity and found little difference between them. 
Since the star formation rate is varied among our simulations, the SB99 spectral energy distributions is modified as,

\begin{equation}
\label{eqn:Loutflow}
L_{\rm outflow} = L_{6,{\rm SB99}}\left(\frac{\dot{M}_{SFR}*t_{\rm evo}}{10^{6} M_\odot}\right),
\end{equation}

\noindent
where $\dot{M}_{SFR}$ is the star-formation rate, t$_{\rm evo}$(=4$\times$10$^{4}$yr), is the time over which the spectrum is aged
and the spectral energy distribution from SB99, $L_{6,{\rm SB99}}$, is normalized by a total stellar mass of 10$^{6}$ M$_{\odot}$.
The conversion of SB99 spectral energy distributions with units of erg s$^{-1}$ \AA$^{-1}$ to spectral radiance is given as, 

\begin{equation}
B_{\nu,{\rm SB99}} = L_{\rm outflow}\left(\frac{\lambda^2}{c}\right)\left(\frac{1}{16\pi^2 R^2_*} \right), 
\end{equation}

\noindent
where $\lambda$ is the wavelength, $c$ is the speed of light and R$_*$ is the sonic radius. 
Combining both the converted SB99 spectra and the meta-galactic background then creates the total composite spectrum. 
Figure~\ref{fig:Spectrum} shows both components of the incident spectra as well as the set of composite spectra used in our simulations. 
As expected the star formation rate scales the spectra linearly in intensity. 
Overall, the meta-galactic background has a much lower intensity compared to the starburst background up to an energy of $\sim$100 eV.
Although we use two components for the background spectra, we find that the photoionization and photoheating rates are dominated by the SB99 spectra and the meta-galactic background plays only a minor role.

The total background is spatially varied and follows an inverse square profile, 
\begin{equation}
\label{eqn:Bnu}
B_{\nu}(R)= \left(B_{\nu,{\rm SB99}}+B_{\nu,{\rm HM}}\right)\left(\frac{R_*}{R}\right)^2,
\end{equation}

\noindent
where R$_{*}$ is the sonic radius, B$_{\nu,{\rm SB99}}$ is the Starburst99 component, and B$_{\nu,HM}$ is the metagalactic background component. 
While the UV background does vary spatially, it is fixed in time. 
A minimum UV background is imposed that corresponds to a strength expected from the meta-galactic flux. 

\subsection{Simulation Setup and Nomenclature}

Each simulation is run in one-dimensional spherical coordinates with the inner radius equal to the sonic radius, which is set to $R_*$=300 pc, consistent with  M82 \citep[\eg][]{Mckeith1995,Strickland2009}. The maximum radius is 10 kpc.  
The ambient medium is set to an initial density that is much lower than the outflow density, $\rho_{\rm ambient}=\rho_{\rm outflow}/1000$ with an initial temperature of $T_{\rm ambient}=10^{6}$ K. 
The metallicity of the ambient medium is set to be equal to the outflow metallicity with ionization states set to their CIE values.
These CIE values are computed assuming a negligible ionizing background.
The actual values for the ambient medium are largely defined for completeness as the density of
the ambient gas is low enough that it is removed by the outflow. 
The base grid of each simulation is comprised of 256 blocks. 
We allow up to four levels of refinement based on the density, pressure, and the radial velocity. This gives a maximum resolution of 2.38 pc. 
Each simulation is run until it reaches steady state, in which $v$, $\rho$, and $T$ remain constant. 

\begin{figure}[!t]
\begin{center}
\includegraphics[trim=0.0mm 0.0mm 0.0mm 0.0mm, clip, width=\columnwidth]{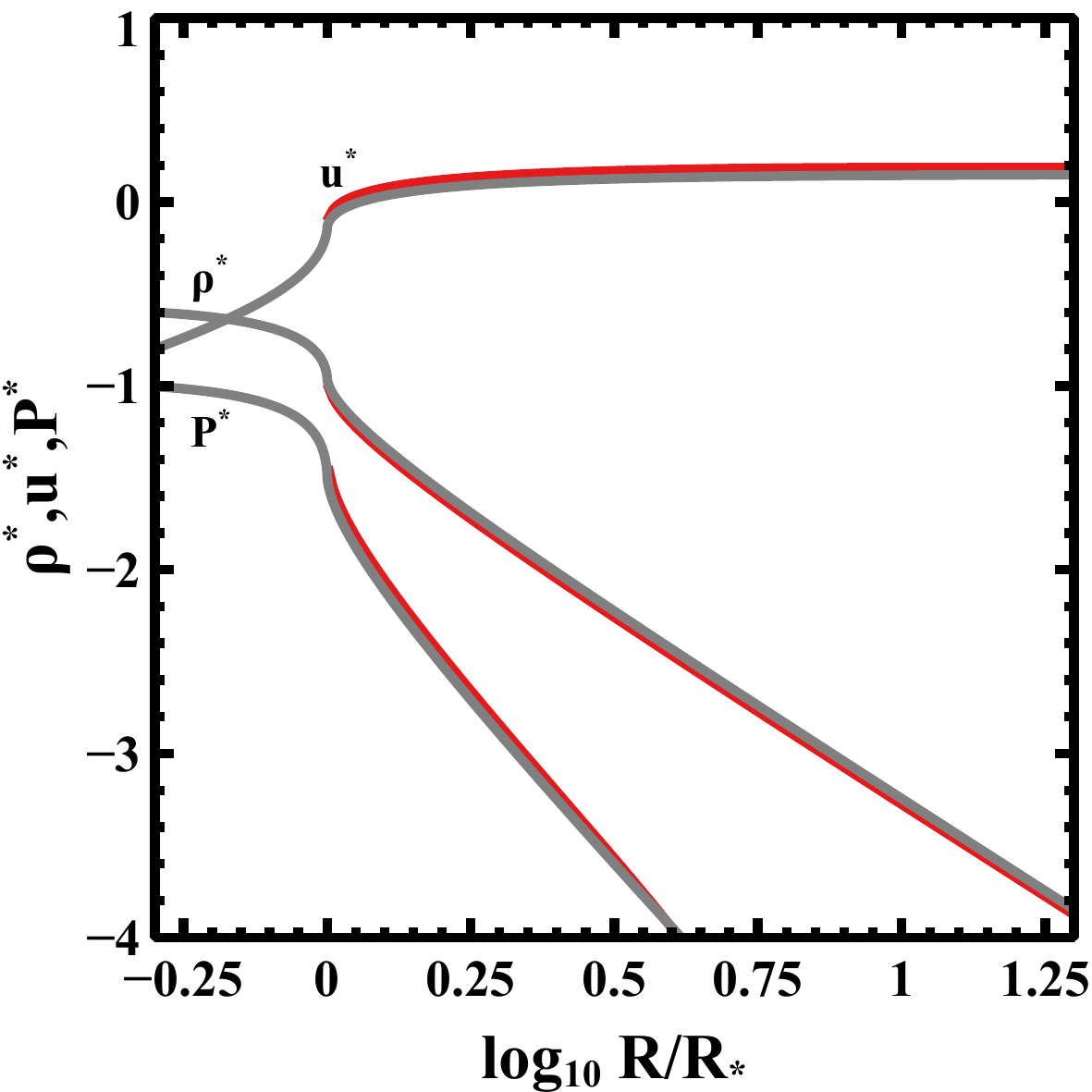}
\caption{Comparison between CC85 (gray lines) solution and our fiducial model (red lines). Although only a single model is shown identical results are found for all models.  }
\label{fig:CCComp}
\end{center}
\end{figure}

Table~\ref{tab:runsummary} summarizes the simulations presented here. 
The mass inflow rate, the energy inflow rate, the sonic radius,  metallicity, and whether or not the UV background is on are what define each simulation. 
The fiducial model is defined as a model with a mass inflow rate of $\dot{M}$=10 M$_{\odot}$~yr$^{-1}$, an energy inflow rate of $\dot{E}$=10$^{50}$ erg~yr$^{-1}$, and solar metallicity.
For simplicity, each model is named by what is varied relative to our fiducial model. 
For example, Z05 represents a model with a mass inflow rate of $\dot{M}$=10 M$_{\odot}$~yr$^{-1}$, an energy inflow rate of $\dot{E}$=10$^{50}$ erg~yr$^{-1}$, and a gas metallicity of half that of the solar metallicity, and with the background UV field. The name is therefore Z05 because only the metallicity is changed relative to the fiducial model.

\section{Results}

\begin{figure}[!t]
\begin{center}
\includegraphics[trim=0.0mm 0.0mm 0.0mm 0.0mm, clip, width=\columnwidth]{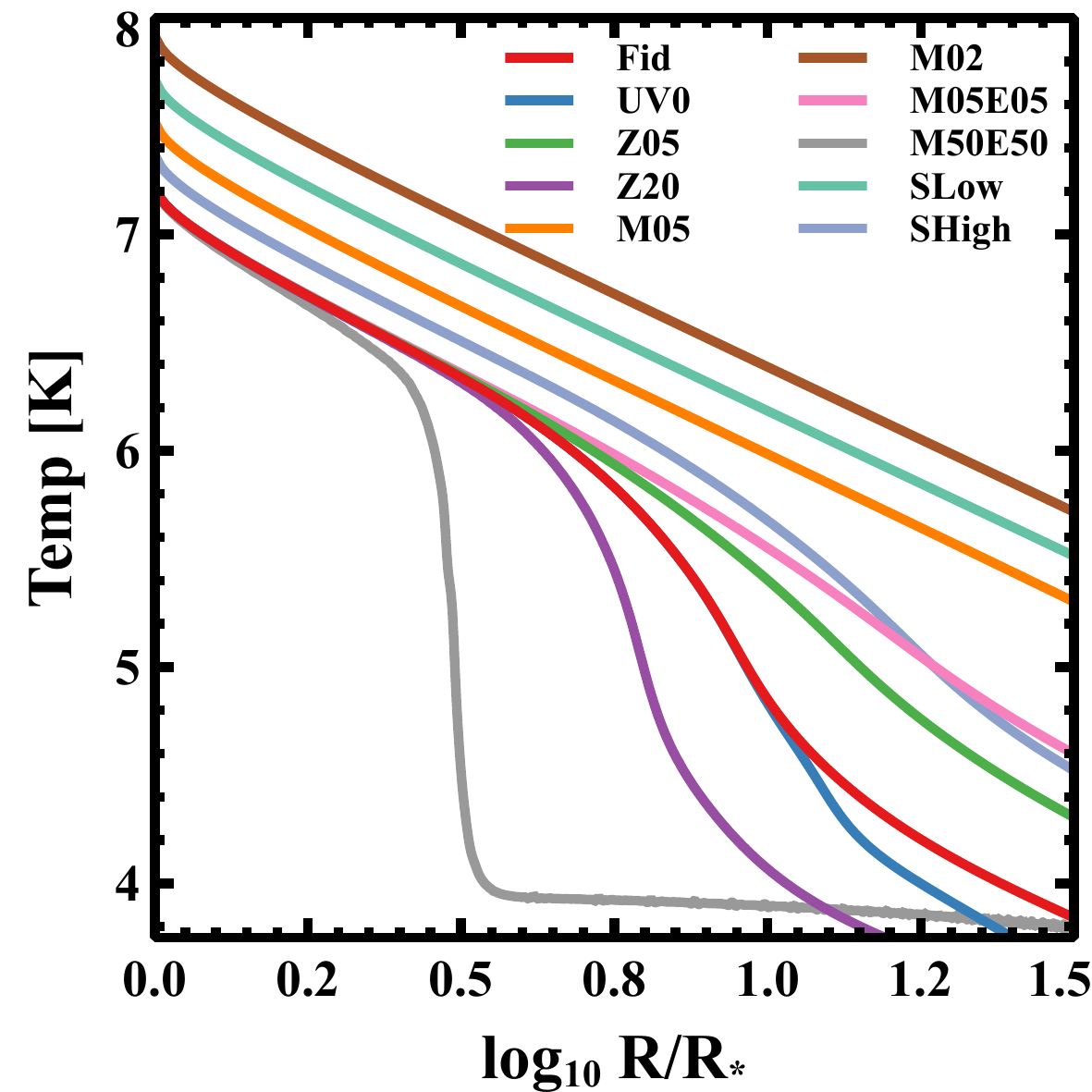}
\caption{Comparison of gas temperatures as a function of radius for all models. The legend at top right provides the line color for each model.}
\label{fig:ModelTemps}
\end{center}
\end{figure}

We show  the normalized density, pressure, and radial velocity for our fiducial model and the CC85 results (Figure~\ref{fig:CCComp}). 
The CC85 results are obtained by using their equations (4) and (5) with the parameters of the fiducial model. 
The sound speed is given by, $c^2_{s}[M^2/2+1/(\gamma-1)]=q_e/q_m$, the 
outflow velocity is then $c_s M=u$, and the density is given by the constraint, $\rho u r^2 = constant$. 
Following CC85, the density, pressure, and radial velocity are normalized by, $\rho^*=\rho/(\dot{M}^{3/2}\dot{E}^{-1/2}R_*^{-2})$, $P^{*}=P/(\dot{M}^{1/2}\dot{E}^{1/2}R_*^{-2})$, and $u^{*}=u/(\dot{M}^{-1/2}\dot{E}^{1/2})$ respectively. 

We find very good agreement between the CC85 solution and our results. 
Note that since the inner radius of our simulations are placed at the sonic point, all of our results start at \logr=0. 
This comparison verifies that our inflow boundary conditions are accurate and that the density of our ambient medium is low enough that is does not substantially impede the outflow. 
For completeness, we show the gas temperature as a function of radius for each model (Figure~\ref{fig:ModelTemps}). 
The initial temperature is given by equation~(\ref{eqn:temp}).

\subsection{Ionization state of the gas in the fiducial case}

\begin{figure}[!t]
\begin{center}
\includegraphics[trim=0.0mm 0.0mm 0.0mm 0.0mm, clip, width=\columnwidth]{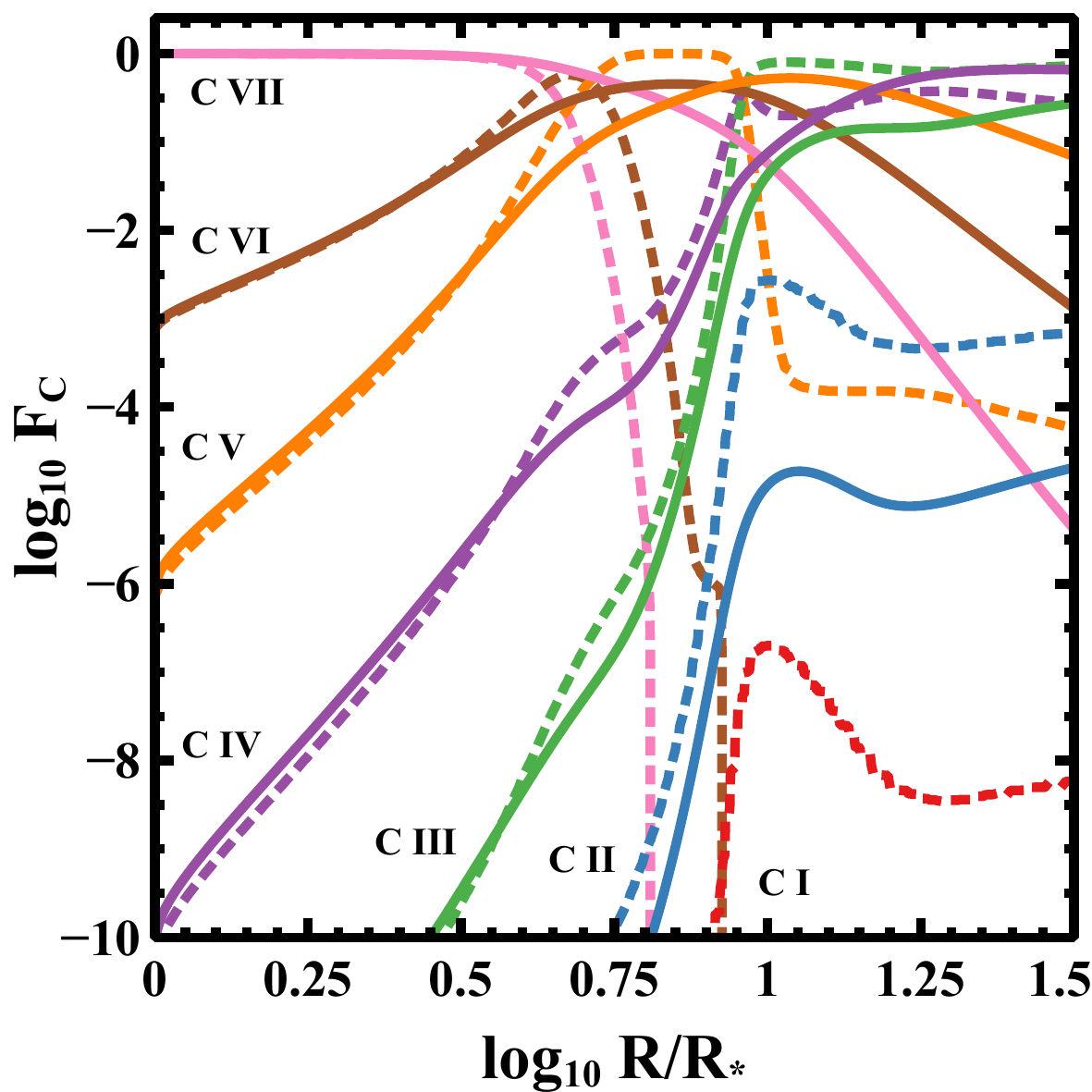}
\caption{Ionization states of carbon for the fiducial model at the end of the simulation. The solid lines show the \flash\ results while the dotted line show the expected values from \cloudy. The ordinate is the fractional ionization state of carbon defined as F$_{i}$ = X$_{i}$/$\sum$ X$_{i}$, where X$_i$ is the mass fraction of ionization state $i$, and the sum is over all ionization states of carbon. The symbols corresponding to each ionization state are identified next to the appropriate line.}
\label{fig:CarbonCIE}
\end{center}
\end{figure}

The utility of tracking the ionization state of individual ions is that we get an accurate
picture of the evolution of the ionization state of the ensemble of elements in the outflow.
Figure~\ref{fig:CarbonCIE} shows the results for all ionization states of carbon.
Here we compare the \maihem\ results with equilibrium results computed using \cloudy\ \citep{Ferland2013}, using our input ionizing background. Initially, Carbon is predominately fully ionized (\ie\ \CVII) and remains in a higher ionization state with respect to the equilibrium ionization at all radii (it is ``over-ionized'').  Near the gas injection boundary at the sonic radius, the densities are the highest, and reactions proceed quickly. Thus the relatively over-ionization is low. In this region, the equilibrium and full-chemistry results follow each other relatively closely.   At larger radii, however, the recombination rates drop and non-equilibrium effects become more pronounced, often differing by orders of magnitude at distances above $\approx$10 times the sonic radius.

Note also that these effects are often difficult to discern given the partial information that is available from optical-UV spectroscopy of galaxies exhibiting outflows. For example at $R/R_\star \approx 10$, and \CIV, \CIII, \CII\ are all lower than their equilibrium values, and could be interpreted as corresponding to lower temperatures and/or lower ionization parameters without the additional information that the most of the carbon is in the form of \CV\ and \CIV\ rather than \CI.

\begin{figure}[!t]
\begin{center}
\includegraphics[trim=0.0mm 0.0mm 0.0mm 0.0mm, clip, width=0.98\columnwidth]{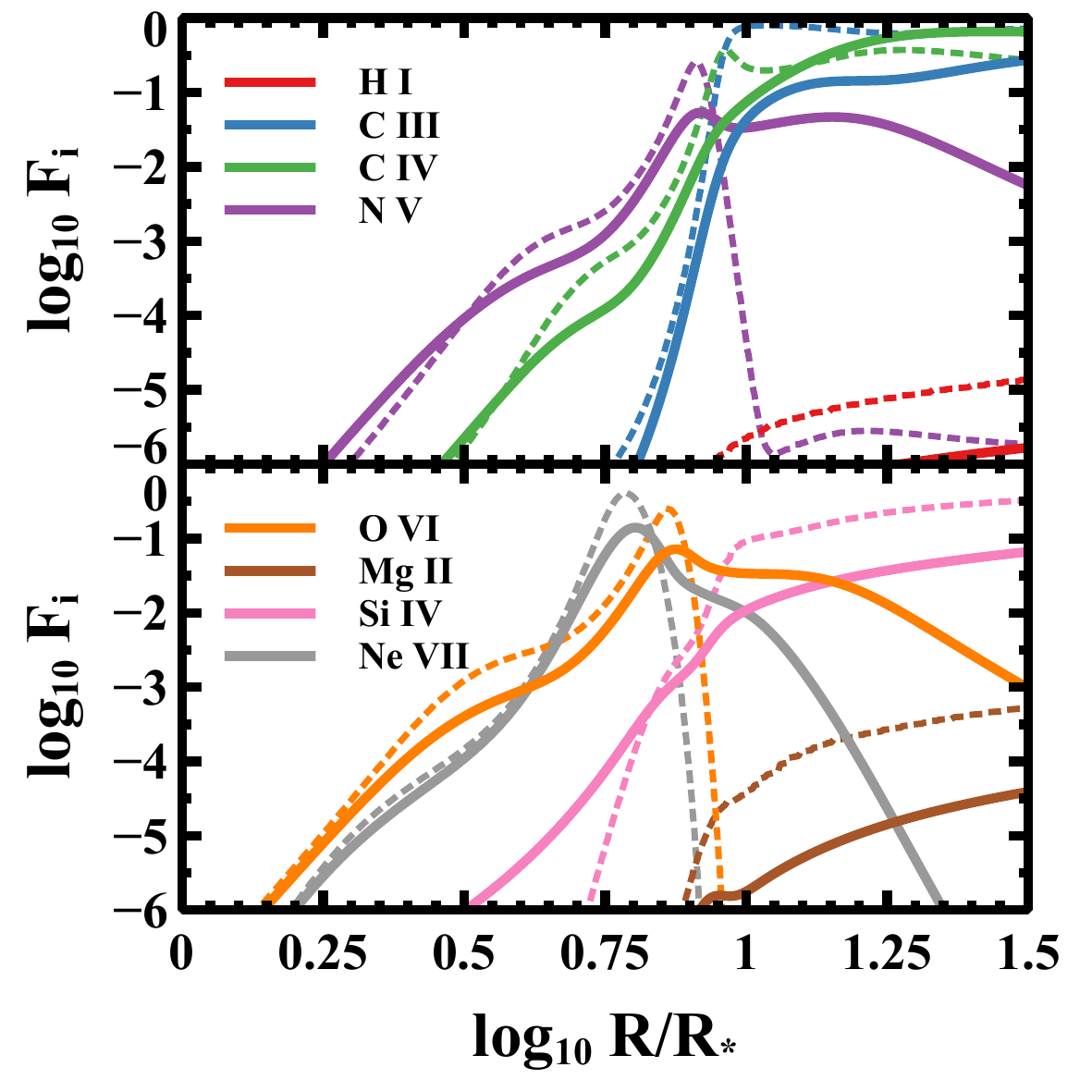}
\caption{Fractional ionization states for several observationally important atomic species. The solid lines show the \maihem\ results while the dashed lines show the equilibrium results computed with \cloudy. The legend give the species and ionization state. The fractional ionization state is defined as F$_{i}$ = X$_{i}$/$\sum$ X$_{i}$, where X$_i$ is the mass fraction of species $i$, and the sum is over all ionization states for a given element.  }
\label{fig:Cloudy}
\end{center}
\end{figure}

\begin{figure}[!t]
\begin{center}
\includegraphics[trim=0.0mm 0.0mm 0.0mm 0.0mm, clip, width=0.48\textwidth]{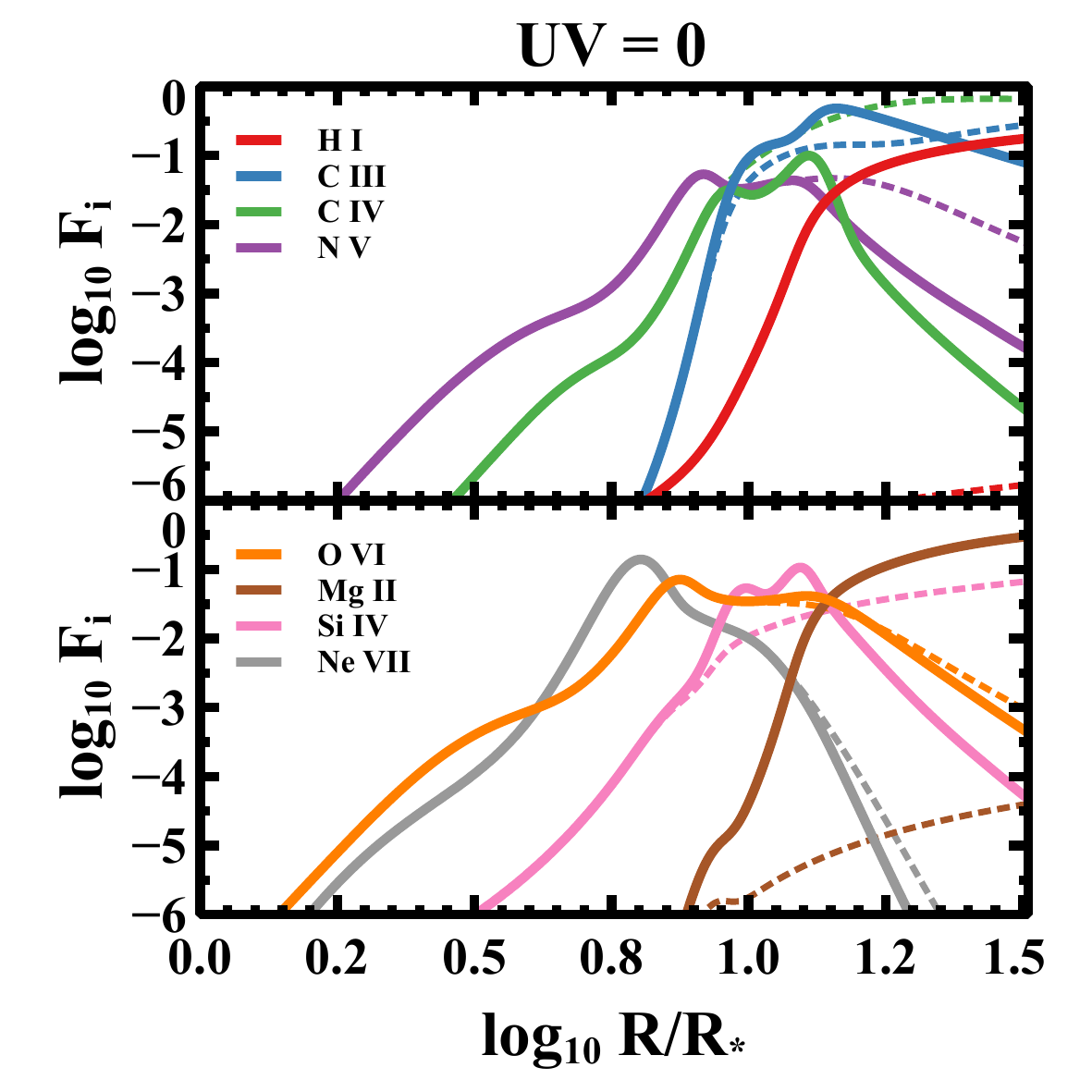}
\caption{ Fractional ionization states for UV0. The solid lines show the results from the model with the parameters given by the title while the dashed lines represent the fiducial model. The legends at the top left of each panel provided the line color associated with each ionization state.  }
\label{fig:UVComp}
\end{center}
\end{figure}

In  Figure~\ref{fig:Cloudy} we show the profile in atomic ionization states that are commonly observed. 
As the carbon figure, 
up to \logr$<$0.5, the outflow ionization states are well described by the equilibrium values. 
At larger radii,  the full \maihem\ calculation  generally produces higher ionization states than those predicted assuming collisional ionization equilibrium, but the fraction of a particular ion of an element is difficult to predict \textit{a priori}.  For example,
 \NV, \OVI, and \NEVIII\ all exceed their equilibrium ionization fractions and are over-ionized, while \CIV, \SiIV, and \MGII\ are much less over-ionized.
These strong differences highlight the importance of considering non-equilibrium ionization when interpreting emission and absorption line spectra of galactic outflows and how outflows may cool and ultimately evolve. 

\subsection{The impact of varying the metallicity and UV background}

We demonstrate the impact of varying the UV background on the ionization state of the outflowing gas in Figure~\ref{fig:UVComp}. 
Only at large distances from the sonic radius, \ie\ \logr$>$1, do some ionization states in the UV-free case deviate from the fiducial case. 
Higher ionization states, such as \NEVII, \SiIV, and \NV, show little difference between the fiducial and UV-free case and only for large radii. 
Lower ionization states, such as \HI\ and \MGII, show large differences.
These lower ionization states have larger photoionization cross sections, and therefore photoionization leads to higher ionization states.
In the absence of a UV background field these low ionization states remain populated compared to the fiducial model.

\begin{figure*}
\begin{center}
\includegraphics[trim=0.0mm 0.0mm 0.0mm 0.0mm, clip, width=0.93\textwidth]{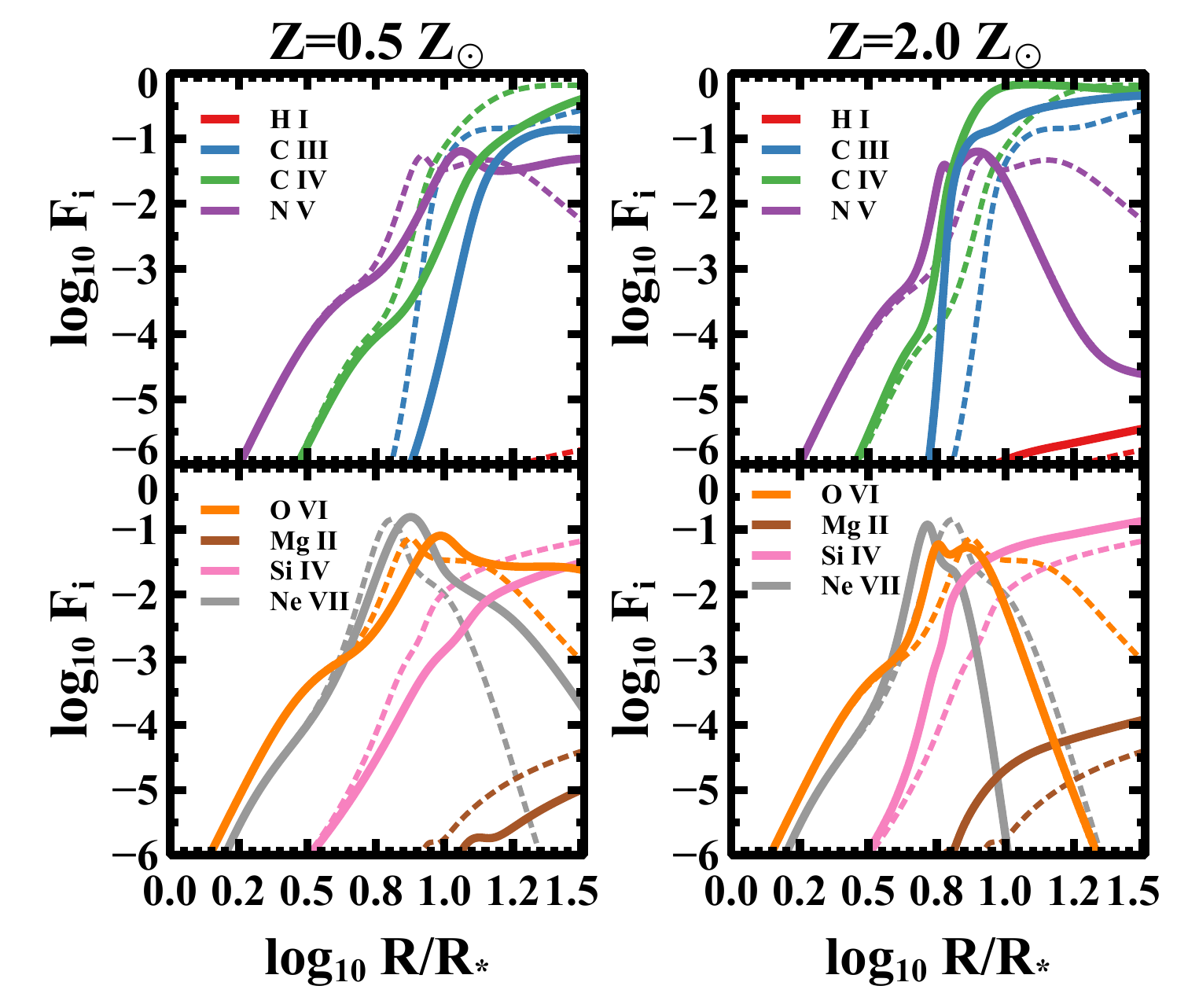}
\caption{ Fractional ionization states for {\it left panel: } Z05 and {\it right panel: } Z20. The solid lines show the results from the model with the parameters given by the title while the dashed lines represent the fiducial model. The legends at the top left of each panel provided the line color associated with each ionization state.  }
\label{fig:ZComp}
\end{center}
\end{figure*}

A range of metallicities are found within galactic outflows. Numerical studies have found that these outflows can enrich the gas within galaxy halos up to supersolar metallicities \citep{Choi2017} while others find the outflows are enriched up to 8Z$_{\odot}$ \citep{Gan2018}. Using a sample of quasars \cite{Xu2018} found that gas within the broad line region to be highly enriched. 
Conversely, outflows with low metallicities have been found in simulations of galaxy formation that employ realistic feedback perscriptions \citep[\eg][]{Hafen2017,Muratov2017}.

Our fiducial model assumes that the gas in the outflow has solar metallicity which we compare to a model with half solar metallicity and twice solar metallicity (Figure~\ref{fig:ZComp}). 
Close to the sonic radius changing the metallicity has little effect on the resulting ionization state. 
Only at large distances from the sonic radius, \logr$>$0.75 do metallicity effects begin to impact the ionization state of the gas. 
In particular, lower metallicity leads to generally higher ionization states of certain elements compared to the fiducial model, \eg \CIII\ and \CIV. Other elements, the reverse is true. Ionization states that are particularly impacted by lowering the metallicity having higher relative ionization are \CIII, \CIV, \NV, and \SiIV.

In general, increasing the metallicity has the effect of creating a more neutral gas. For example, the \NEVIII\ and \OVI\ curves show that these ions peak sooner and fall to much lower abundances at larger radii compared to the fiducial case. Others, such as \SiIV\ and \MGII\ have similar profiles as the fiducial model, but have overall higher abundances.
The impact of metallicity on altering the ionization state of the gas is due to the changes in recombination time.
The recombination time scale is approximated as $\tau_{\rm rec}=1/Rn_e$ where $R$ is the total recombination rate which is a strong function of temperature and n$_e$ is the electron density.
In general the recombination rate is smaller for hotter temperatures.
Therefore, the recombination times are longer for Z05 because there are fewer electrons and the gas remains hotter.
Z20, on the other hand, has shorter recombination times due to the increase in the electron densities and overall cooler gas temperatures, as shown in Figure~\ref{fig:ModelTemps}.

\subsection{The impact of varying the mass outflow rate}

\begin{figure*}
\begin{center}
\includegraphics[trim=0.0mm 0.0mm 0.0mm 0.0mm, clip, width=\textwidth]{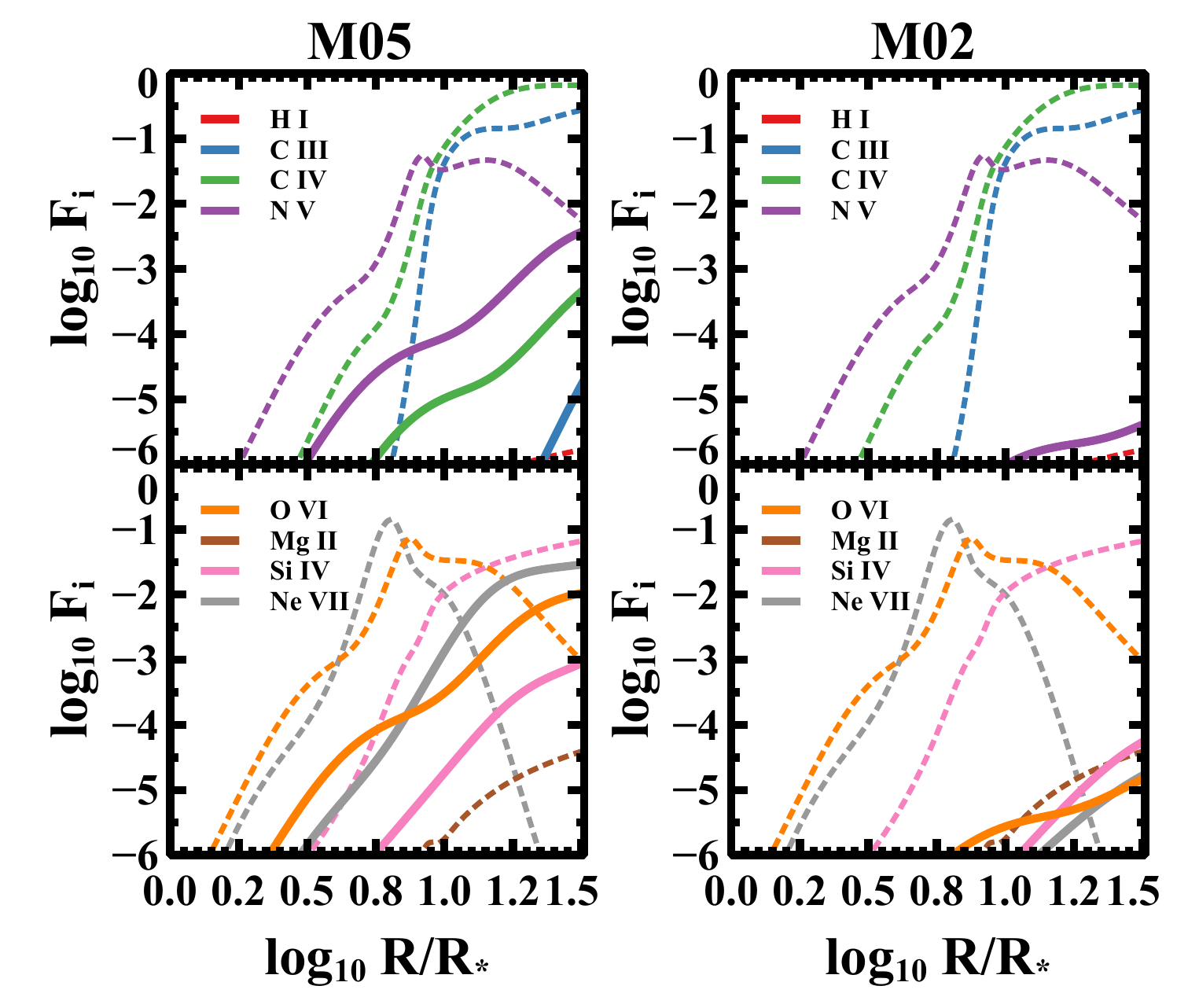}
\caption{Fractional ionization states for two models with $\dot{M}$ = 5 M$_{\odot}$~yr$^{-1}$ ({\it left panel }) and $\dot{M}$=2~M$_{\odot}$~yr$^{-1}$ ({\it right panel}). The energy inflow rate is fixed at $\dot{E}=$10$^{50}$~erg~s$^{-1}$. The solid lines show the results from the model with the parameters given at the top of each upper panel of the figure while the dashed lines represent the fiducial model. The legend indicates the line color used for each ionization state.}
\label{fig:MdotComp}
\end{center}
\end{figure*}

We show the effect of changing the mass inflow rate at the simulation boundary in
Figure~\ref{fig:MdotComp}. 
Two models, M05 and M02, were calculated that decrease the mass inflow rate by 50 and 80\% relative to the fiducial model while keeping the energy inflow rate constant (\ie\ decreasing the relative amount of the mass-loading). 
Decreasing the mass injection rate has the effect of increasing the gas velocity that leads to a decrease in the gas density, increases the gas temperature, and increases the overall ionization state of the outflowing gas.
For example, the relative fraction of the ion \NV\ is reduced by two orders-of-magnitude in the M05 model compared to the fiducial model. 
This reduction in the fractional abundance of each state is seen across all species and is even more pronounced in the M02 model with its much lower mass injection rate.
In fact, for M02 only the highest ionization states are found, such as \NV, \OVI\, and \NEVII\ have any significant fractional abundance. 
The ionization states also have a much smoother profile in M05 and M02 compared to the fiducial model, that is, it does not have any peaks in ionization state found in the fiducial model. 
This is due to the higher initial outflow temperature which never falls below T$\approx$10$^{6}$K, which prevents the recombination to the ionization states we consider from becoming very efficient. 
Thus, the differences in ionization state are a simple consequence of longer recombination times for these atomic species. 
The outflow density in these models is lower than the fiducial model which decreases the number density of electrons, and consequently, increases the recombination time scale and the winds stay highly ionized. 

\subsection{The impact of varying the flow density}

\begin{figure*}
\begin{center}
\includegraphics[trim=0.0mm 0.0mm 0.0mm 0.0mm, clip, width=\textwidth]{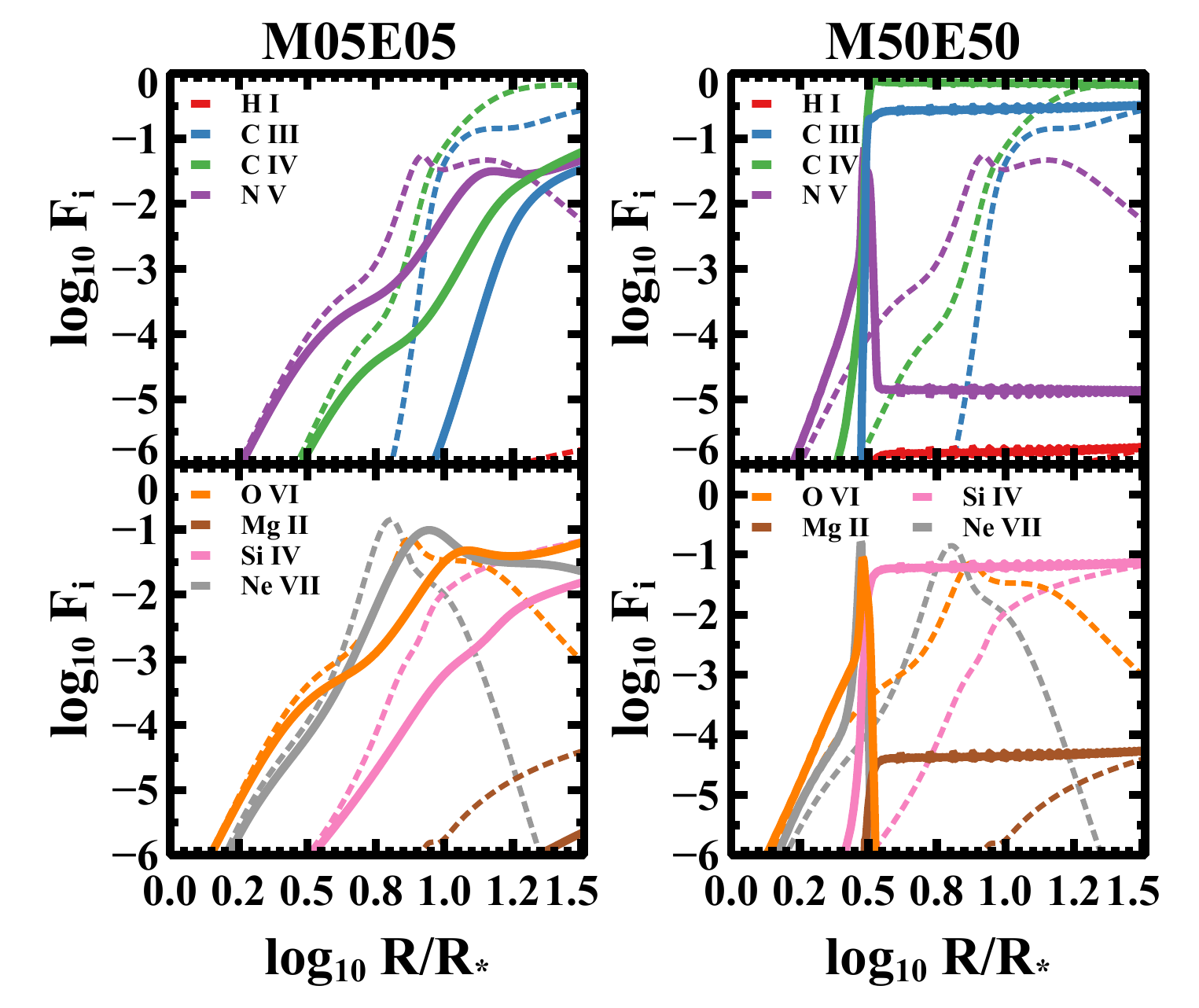}
\caption{Fractional ionization states for the model, M05E05 {\it (left panel)}  and M50E50 {\it (right panel)}. The solid lines show the results from the model with the parameters enumerated on the top of each panel and the dashed lines represent the ionization fraction of the fiducial model. The legend at the top of each panel indicates the line color used for each ionization state.}
\label{fig:MassFluxComp}
\end{center}
\end{figure*}

We next analyze the impact of varying both the mass and energy outflow rates while keeping the outflow velocity fixed (Table~\ref{tab:runsummary}). 
This leads to outflows with varying initial mass densities but the same initial temperatures.
Two models are run that vary both outflow rates (or inflow rates into the simulation). The model, M05E05, decreases both rates by a factor of two while the model, M50E50, increases them by a factor of five. 
This has the effect of raising or lowering the outflowing gas density while keeping the initial gas outflow velocity and temperature constant. 

One additional change is made for M50E50 regarding the star formation rate. 
We note that these models are similar to those presented in \cite{Scannapieco2017}, where M05E05 represents conditions similar to the starburst galaxy, M82 \citep[\eg][]{Forester2003}, while M50E50 represents conditions seen in ultra-luminous infrared galaxies \citep[ULIRGs;][]{Daddi2005,Piqueras2012}. 
In order to achieve such high mass outflow rates, the star-formation rate within the sonic radius must also be effectively increased. 
For M50E50, our input parameters are equivalent to assuming that the star-formation rate is a factor of ten higher than that of the M05E05 model. 
Due to the way we scale the UV background, increasing the star-formation rate also increasing the intensity of the ionizing radiation field leading to higher photoionization and photo-heating rates.

We show the results of these two models in Figure~\ref{fig:MassFluxComp}. 
At small radii, \logr $<$0.5, M05E05 matches very well with the fiducial model for the ionization states shown. 
At larger radii, M05E05 shows lower ionization mass fractions compared to the fiducial model.
For example, both \NV\ and \CIV\ have peaks in their ionization that are slightly downwind of the fiducial model. 
Similar structure is seen in \OVI\ and \NEVII\ but is more pronounced.
In terms of the temperature, M05E05 is hotter at larger radii when compared to the fiducial model. 

M50E50 shows much more dramatic differences with respect to the fiducial model than M05E05. 
In this model, the gas density is high enough that cooling and recombination become very efficient. 
In fact, at \logr$=0.5$ the gas temperature quickly decreases to T$\approx$10$^{4}$ K and leads to a unique distribution of ionization states where many of the higher ionization states quickly recombine (\eg\ \NV\ and \NEVIII) leading to high contributions to their elemental mass fractions for the ions, \CIII\ and \SiIV. Interestingly, the contribution to the overall ionization of O from \OVI\ is negligible for radii beyond \logr $\approx$0.5.
At larger radii the ionization states remain largely unchanged, except for \NV, suggesting that equilibrium values are reached. 
This equilibrium value is found in balancing the electron recombination rate to the photoionization rate. 
At T$\approx$10$^{4}$ K, the equilibrium temperature, electron impact ionization rates are typically small compared to either the electron recombination rates or the photoionization rates. 
As mentioned above, the recombination rate scales as the inverse of the electron number density and therefore has an $r^{-2}$ spatial profile, that is, the recombination timescale increases with increasing distance from the starburst. 
Similarly, the photoionization rate goes as $\Gamma_{i}=1/\gamma_{0,i}J_{21}$, where $\gamma_{0,i}$ is the normalized photoionization rate for ion $i$ and J$_{21}$ quantifies the strength of the UV background. 
As such, it follows eq.~(\ref{eqn:Bnu}) and has an $r^{-2}$ spatial profile. 
Therefore, both the recombination and photoionization time scales follow the same $r^{-2}$ profile and the ionization states are in equilibrium for \logr$>$0.5.

We also note that M50E50 may represent a case of catastrophic cooling within the outflow and may provide important insight into the development and evolution of superwinds and wind-driven superbubbles \cite[\eg][]{Silich2007,Silich2017}. 
In particular, it may prove useful in understanding feedback processes in young super star clusters such as Mrk 71 \cite[\eg][]{Oey2017}. 
The interaction of the outflow with a hydrodynamically important ambient medium is outside the scope of this work and will form the basis of a future study.

\subsection{Comparison to \cite{Schneider2018A}}

\begin{figure*}
\begin{center}
\includegraphics[trim=0.0mm 0.0mm 0.0mm 0.0mm, clip, width=\textwidth]{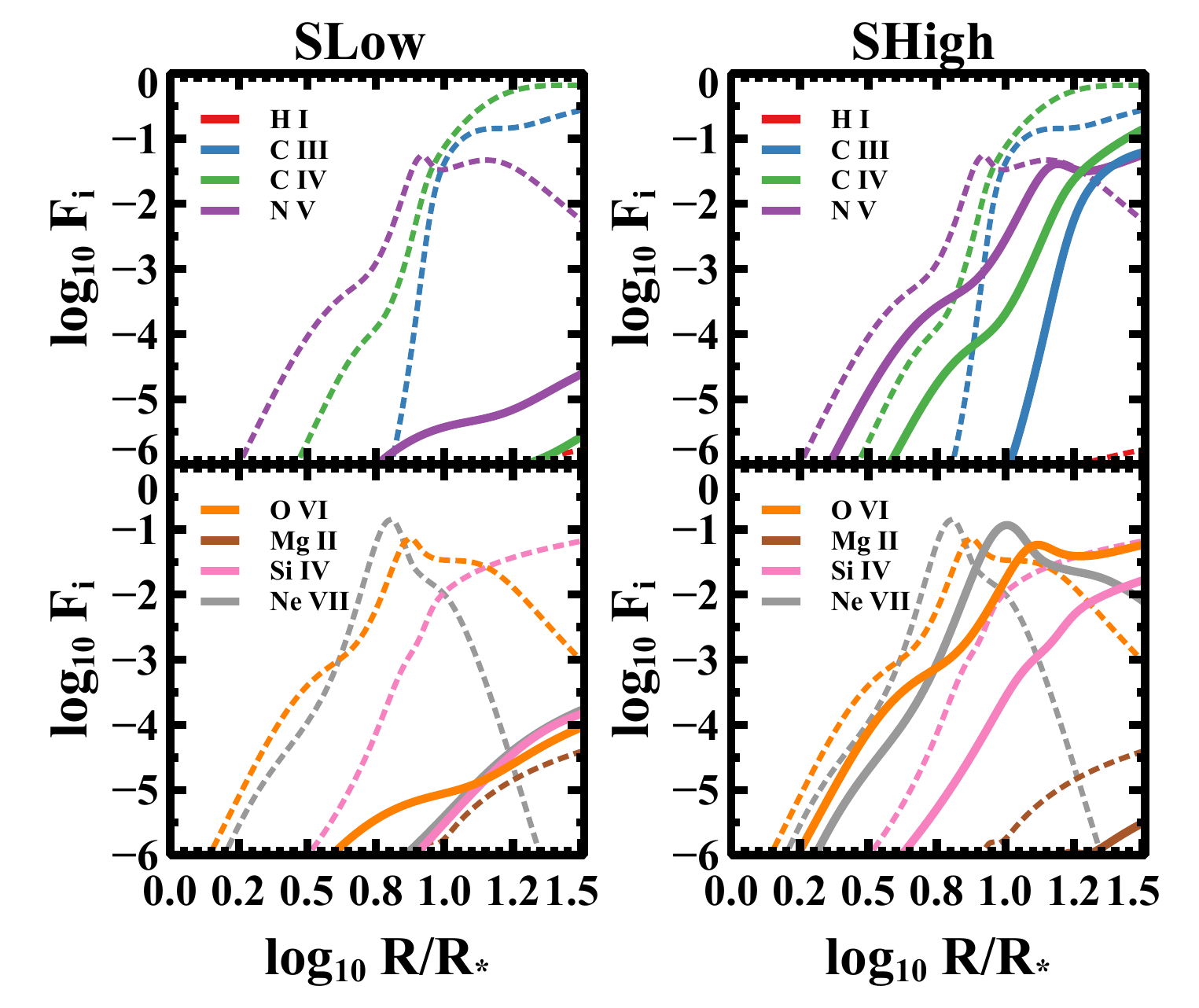}
\caption{Fractional ionization states for a low mass-loading, SLow ({\it left panel}), and a high mass-loading model, SHigh ({\it right panel}).
See Table~\ref{tab:runsummary} and the text for the details of the input parameters of these two models. The solid lines show the results from the model with the parameters given at the top of the panels and the dashed lines represent the results of our fiducial model. The legend in each panel indicates the line color used for each of the ionization states.} \label{fig:SchneiderComp}
\end{center}

\end{figure*}
In the high-resolution galaxy outflow simulations presented in \cite{Schneider2018A}, the CC85 model is used as the basis for their outflow model.

Two outflow conditions are considered by these authors, a ``high mass-loading'' (SHigh) model with a mass and energy injection rate of $\dot{M}$=12~M$_{\odot}$~yr$^{-1}$ and $\dot{E}$=1.7$\times$10$^{50}$~erg~yr$^{-1}$; and a ``low mass-loading'' (SLow) model with rates of $\dot{M}$=1.5~M$_{\odot}$~yr$^{-1}$ and $\dot{E}$=0.47$\times$10$^{50}$~erg~yr$^{-1}$. 
As discussed by these authors, both cases represent physically motivated conditions. The ``high-mass loading'' model represents starburst-like conditions early in the burst where the star formation rate is high. 
The ``low-mass loading'' model represents a more mature starburst where the star formation rate is lower and gas in the ISM has been swept away.
These two parameters set then span an interesting range in evolutionary time for a given starburst powered outflow.

We reproduced these two cases using our code (Figure~\ref{fig:SchneiderComp}). 
Compared to our fiducial model, the low mass-loading case produces higher ionization states at all radii. 
In fact, almost all of the studied ionization states are absent; only the highest ionization states are present but in very low abundance.
This is reflected in the temperatures of each model that are much hotter in the simulations using the Figure~\ref{fig:ModelTemps} initial conditions than the fiducial model (Figure~\ref{fig:ModelTemps}). 
These high temperatures create an environment where recombination rates are very low and the ionization state remains largely unchanged as the gas flows outward from the sonic radius.

The high mass-loading case, on the other hand, produces ionization states that are more similar to the fiducial case. 
This model produces peaks in \NV, \NEVII, and \OVI\ ionization states that are slightly higher radii from the sonic radius. 
For example, the peak of \NEVII\ is at \logr$\approx$0.75 in the fiducial model but found at \logr$\approx$1.0 in the high-mass loading case. 
This too is reflected in Figure~\ref{fig:ModelTemps} where the high-mass loading case remains slightly hotter than the fiducial model. 

\section{The column densities of \NV\ and \OVI}
\begin{figure}
\begin{center}
\includegraphics[trim=0.0mm 0.0mm 0.0mm 0.0mm, clip, width=0.95\columnwidth]{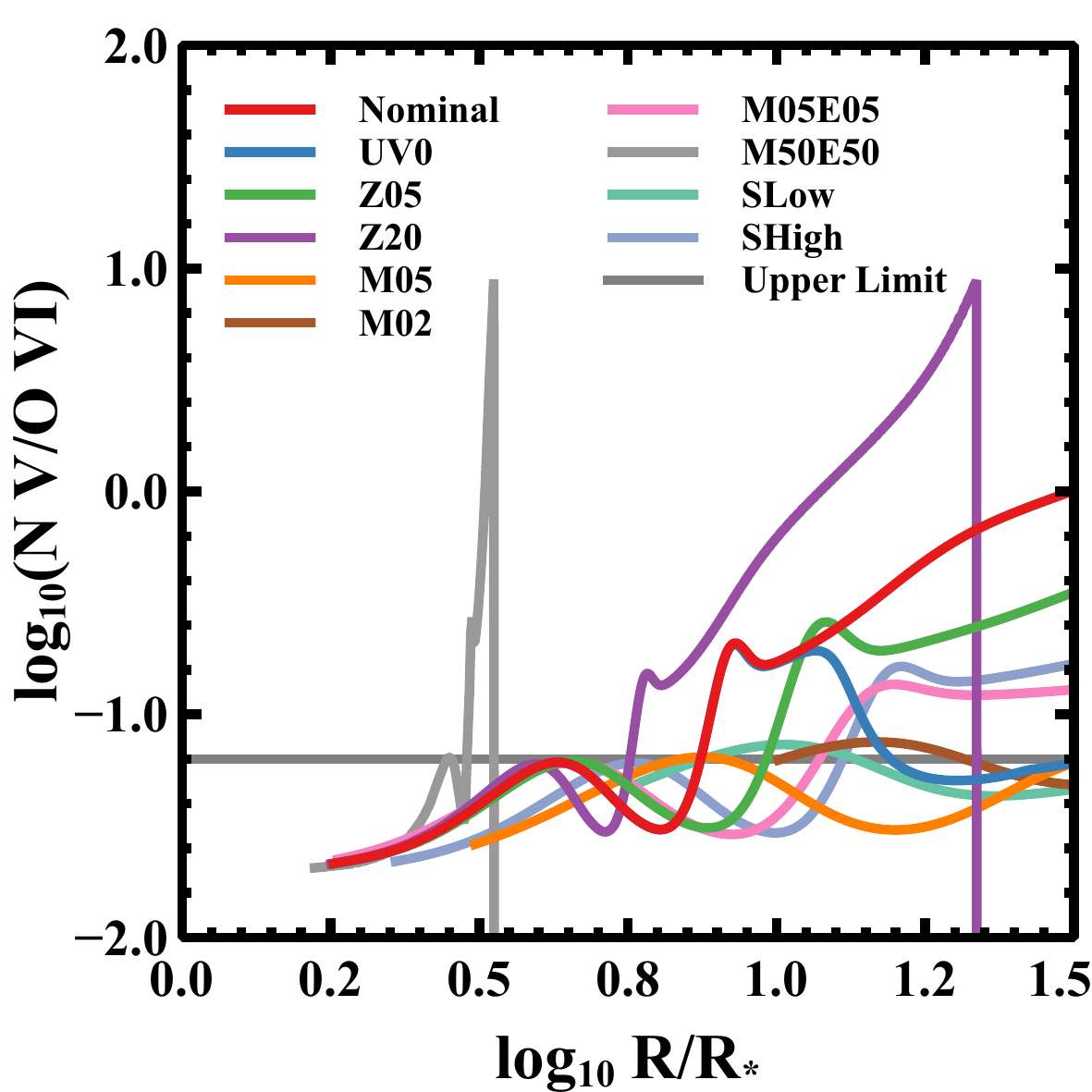}
\caption{ \NV-\OVI\ column density ratios. Each line represents a different model as indicated in the legend at the top of the panel. The gray solid line shows the upper limit on the column density ratio of \NV\ and \OVI\ from \cite{Chisholm2018}. }
\label{fig:OVINV}
\end{center}
\end{figure}

X-ray observations of galactic outflows show that the hot gas (T$>$10$^{7}$ K) dominates the kinetic and thermal energy of the outflow \citep[\eg][]{Martin1999,Griffiths2000,Strickland2000,Strickland2009,Zhang2014}. 
Since the hottest gas in outflows, $\sim$10$^8$ K, \ie\ the ``piston'' that is driving the wind, is diffuse and has low emissivity, observations of this phase are particularly difficult. The best way to study the piston is to observe emission lines in the X-ray that are sensitive to gas at such high temperatures \citep{Strickland2009}.  Alternatively, one can study the strong UV/optical emission lines to probe the cooling rate and kinematics of outflowing gas at intermediate temperatures, 10$^{4}$ to 10$^{5.5}$ K \citep[\eg][]{Lehnert96,Hayes2016}. More commonly however, UV and optical absorption lines are used to probe the terminal velocities, column densities, covering fractions, and energetics of the outflowing gas over these intermediate temperatures \citep[\eg][]{Heckman2015, Heckman2017, Chisholm2018}.

\cite{Chisholm2018} used the \OVI\ 1032 \AA\ doublet in the rest-frame far UV to study the column densities of warm/hot gas at
temperatures $\approx$10$^{5.5}$ K and cooler. The authors used photoionization models to predict the abundance of \OVI\ finding that photoionization alone was unable to explain the \OVI\ column densities. 
But importantly for testing models of the gas ionization, they also found that \NV\ is not detected in absorption even though \NV\ 1243 \AA\ is expected to be a relatively strong transition and probe similar gas. 
They found an upper limit on the \NV-\OVI\ column density ratio of $\approx$0.07. 

As these photoionization models assume photoionization equilibrium, we are interested in the column densities of \NV\ and \OVI\ and their ratio produced by the suite of models presented here. The total column densities of our expanding wind models are mostly between about 7-24$\times$10$^{14}$ and 7-33$\times$10$^{13}$ cm$^{-2}$ for \OVI\ and \NV\ respectively.  There are models with significantly lower column densities that are always those with very low mass outflow rates.
Our estimated column densities are in good agreement with observations of nearby starburst and gas in the inner halos of galaxies \citep{Grimes2009, Hayes2016, Chisholm2016, Bordoloi2016}.

Figure~\ref{fig:OVINV} shows the ratio of the number densities of \NV\ and \OVI\ for each model. 
In order to prevent numerically unphysical ratios, only outflow regions where both species are found in appreciable amounts are considered as they are the regions that significantly contribute to the total column of these ions. 
A cutoff value of F$_{i}>$10$^{-6}$ is used for the fractional ionization state of both ions (see the caption of Figure~\ref{fig:Cloudy}). 
At radii of \logr$<$0.75 nearly every model predicts the low abundance of \NV\ compared to \OVI. 
Beyond \logr$>$0.75, the presence of \NV\ is model dependent. 
For example, the fiducial models shows the column density ratio of \NV\ to \OVI\ peaking above the upper limit of \cite{Chisholm2018} at \logr$>$0.9.
Models with lower outflow densities and higher outflow velocities (and outflow temperatures), have ratios that consistent with this upper-limit, \eg\ M05, M02, and SLow.
The lack of a detection of a significant column of \NV\ is a consequence of non-equilibrium effects and does not require alternative mechanisms of generating excess \OVI. However, models with low densities generally do not agree with column densities observed in starbursts, as they are too low.

A plausible way that the models can be reconciled with observations, namely, producing sufficient \OVI\ column densities while preserving a low column density ratio of \NV\ and \OVI\ is to increase the terminal velocity of the wind for constant mass outflow rate and/or have a moderately low outflow rate at constant wind velocity (models that appear roughly analogous to M82 - M05E05).  In other words, within the context of our model, the mass loading of the wind is relatively modest.  Even though some of the models are close to the upper limit observed in the column density (0.1 or less), a more complete set of models with more detailed physics and a wider range of observations of this ratio in starburst galaxies are necessary to see if this is quantitatively correct.

\section{Summary and Conclusion}

We have presented a suite of one-dimensional galaxy outflow models that track the non-equilibrium evolution of the ionization fractions of many species as a function of the sonic radius. 
The outflow model is based on the outflow model of \cite{Chevalier1985} and is implemented as a boundary condition. 
By using the \cite{Chevalier1985} model, the outflow is defined by three parameters at the boundary of the simulation, the mass inflow rate, $\dot{M}$, the energy inflow rate, $\dot{E}$, and the sonic radius, $R_{*}$.  
The inflowing gas is assumed to be in collisional ionization equilibrium as it enters the computational domain. 
In addition, we implement a spatially varying but constant photoionizing background based on Starburst99 models \citep{Leitherer1999} and the meta-galactic UV background at z=0 \citep{Haardt2012}. 

A range of models is presented that vary several parameters, including $\dot{M}$, $\dot{E}$, the metallicity of the gas, and the presence of the UV background. 
We find that every model reaches a steady state hydrodynamic solution that is well described by the normalized \cite{Chevalier1985} profiles of density, pressure, and outflow velocity. 
However, the distribution of ionization states of the species investigated is strongly dependent on the parameters of the outflow. 
Specifically, we find that reactions rates can not keep up with evolution of the outflowing material,  and in many of the cases, the column densities of ions are much higher than one would predict assuming photoionization equilibrium. 

To understand the non-equilibrium effects, we compared the ionization states from our fiducial model to collisional ionization equilibrium models from \cloudy\ \citep{Ferland2013}. 
Two regimes in the ionization are found.
Near the outflows sonic radius, where the wind becomes freely expanding, the model ionization states are well described by equilibrium models.
At a radius of \logr$\approx$0.75, which for the starburst galaxy, M82, is approximately 1.5 kpc \citep[\eg][and references therein]{Strickland2009}, the outflow gas has cooled enough that recombination becomes efficient for the higher ionization states.
At larger radii, \logr$>$0.75, our models are generally have distributions in their ionization states that are skewed to higher ionization states compared to models that assume photoionization equilibrium. This is due to a combination of various non-equilibrium effects and but especially the long recombination times as the density decreases with radius.

One model in particular warrants further study.  M50E50, characterized by large mass and energy inflow rates at the boundary, has the properties of an outflow undergoing catastrophic cooling.  When compared to other models, M50E50 shows a drastic temperature falloff at \logr$\approx$0.6 (see Figure~\ref{fig:ModelTemps}) and an ionization state distribution that seems to be in equilibrium with the ionizing background. Future work will study this phenomenon more closely and its relation to young super clusters \citep{Oey2017} and interaction with a hydrodynamically important ambient medium. 

Finally, we looked at the \NV-\OVI\ column density ratio. 
\cite{Chisholm2018} used the \OVI\ 1032\AA\ far UV absorption line to study the outflow mass loading in a $z\approx$ 2.9 galaxy. 
Interestingly, they were unable to detect the \NV\ 1243\AA\ absorption line even though it should be a prominent feature. 
The authors gave an upper limit of log(\NV/\OVI)$\approx$-1.2 and possible explanations for its absence. 
We have shown here that this ratio can be explained by the non-equilibrium nature of the ionization states within the outflow and is predicted over a wide range of outflow conditions.

The simulations presented here provide a key insight into the ionization state distribution within galactic outflows. 
Although the general structure of the ionization states is similar between models, the exact peaks in the abundance of a given ionization state of an element is model dependent. 
Thus, these models highlight the importance of accounting for non-equilibrium effects in studies of galaxy outflows.

\acknowledgments
We would like to thank Sally Oey and Todd Thompson for useful comments and discussions.
Helpful comments by the referee are also gratefully acknowledged. 
The software used in this work was in part developed by the DOE NNSA-ASC OASCR Flash Center at the University of Chicago. E.S. was supported by NSF grants AST11-03608 and AST14-07835 and NASA theory grant NNX15AK82G. The figures and analysis presented here were created using the {\bf yt } analysis package \citep{Turk2011}. 

\software{FLASH \citep{Fryxell2000}, yt \citep{Turk2011}, CLOUDY\citep{Ferland2013}}
\bibliographystyle{apjsingle}
\bibliography{ms.bib}

\begin{thebibliography}{109}
\expandafter\ifx\csname natexlab\endcsname\relax\def\natexlab#1{#1}\fi
\setlength{\itemsep}{0.0\baselineskip}

\bibitem[{{Agertz} \& {Kravtsov}(2015){Agertz} \& {Kravtsov}}]{Agertz2015}
{Agertz}, O. \& {Kravtsov}, A.~V. 2015, \apj, 804, 18

\bibitem[{{Arribas} {et~al.}(2014){Arribas}, {Colina}, {Bellocchi}, {Maiolino},
  \& {Villar-Mart{\'{\i}}n}}]{Arribas2014}
{Arribas}, S., {Colina}, L., {Bellocchi}, E., {Maiolino}, R., \&
  {Villar-Mart{\'{\i}}n}, M. 2014, \aap, 568, A14

\bibitem[{{Benson} {et~al.}(2003){Benson}, {Bower}, {Frenk}, {Lacey}, {Baugh},
  \& {Cole}}]{Benson2003}
{Benson}, A.~J., {Bower}, R.~G., {Frenk}, C.~S., {Lacey}, C.~G., {Baugh},
  C.~M., \& {Cole}, S. 2003, \apj, 599, 38

\bibitem[{{Bieri} {et~al.}(2016){Bieri}, {Dubois}, {Silk}, {Mamon}, \&
  {Gaibler}}]{Bieri2016}
{Bieri}, R., {Dubois}, Y., {Silk}, J., {Mamon}, G.~A., \& {Gaibler}, V. 2016,
  \mnras, 455, 4166

\bibitem[{{Bolatto} {et~al.}(2013){Bolatto}, {Warren}, {Leroy}, {Walter},
  {Veilleux}, {Ostriker}, {Ott}, {Zwaan}, {Fisher}, {Weiss}, {Rosolowsky}, \&
  {Hodge}}]{Bolatto2013}
{Bolatto}, A.~D., {Warren}, S.~R., {Leroy}, A.~K., {Walter}, F., {Veilleux},
  S., {Ostriker}, E.~C., {Ott}, J., {Zwaan}, M., {Fisher}, D.~B., {Weiss}, A.,
  {Rosolowsky}, E., \& {Hodge}, J. 2013, \nat, 499, 450

\bibitem[{{Bordoloi} {et~al.}(2016){Bordoloi}, {Rigby}, {Tumlinson}, {Bayliss},
  {Sharon}, {Gladders}, \& {Wuyts}}]{Bordoloi2016}
{Bordoloi}, R., {Rigby}, J.~R., {Tumlinson}, J., {Bayliss}, M.~B., {Sharon},
  K., {Gladders}, M.~G., \& {Wuyts}, E. 2016, \mnras, 458, 1891

\bibitem[{{Bordoloi}(2014)}]{Bordoloi2014}
{Bordoloi}, R. e.~a. 2014, \apj, 794, 130

\bibitem[{{Br{\"u}ggen} \& {Scannapieco}(2016){Br{\"u}ggen} \&
  {Scannapieco}}]{Bruggen2016}
{Br{\"u}ggen}, M. \& {Scannapieco}, E. 2016, \apj, 822, 31

\bibitem[{{Chevalier} \& {Clegg}(1985){Chevalier} \& {Clegg}}]{Chevalier1985}
{Chevalier}, R.~A. \& {Clegg}, A.~W. 1985, \nat, 317, 44

\bibitem[{{Chisholm} {et~al.}(2018){Chisholm}, {Bordoloi}, {Rigby}, \&
  {Bayliss}}]{Chisholm2018}
{Chisholm}, J., {Bordoloi}, R., {Rigby}, J.~R., \& {Bayliss}, M. 2018, \mnras,
  474, 1688

\bibitem[{{Chisholm} {et~al.}(2016){Chisholm}, {Tremonti}, {Leitherer}, {Chen},
  \& {Wofford}}]{Chisholm2016}
{Chisholm}, J., {Tremonti}, C.~A., {Leitherer}, C., {Chen}, Y., \& {Wofford},
  A. 2016, \mnras, 457, 3133

\bibitem[{{Choi} {et~al.}(2017){Choi}, {Ostriker}, {Naab}, {Somerville},
  {Hirschmann}, {N{\'u}{\~n}ez}, {Hu}, \& {Oser}}]{Choi2017}
{Choi}, E., {Ostriker}, J.~P., {Naab}, T., {Somerville}, R.~S., {Hirschmann},
  M., {N{\'u}{\~n}ez}, A., {Hu}, C.-Y., \& {Oser}, L. 2017, \apj, 844, 31

\bibitem[{{Cole} {et~al.}(2000){Cole}, {Lacey}, {Baugh}, \& {Frenk}}]{Cole2000}
{Cole}, S., {Lacey}, C.~G., {Baugh}, C.~M., \& {Frenk}, C.~S. 2000, \mnras,
  319, 168

\bibitem[{{Creasey} {et~al.}(2013){Creasey}, {Theuns}, \&
  {Bower}}]{Creasey2013}
{Creasey}, P., {Theuns}, T., \& {Bower}, R.~G. 2013, \mnras, 429, 1922

\bibitem[{{Daddi} {et~al.}(2005){Daddi}, {Dickinson}, {Chary}, {Pope},
  {Morrison}, {Alexander}, {Bauer}, {Brandt}, {Giavalisco}, {Ferguson}, {Lee},
  {Lehmer}, {Papovich}, \& {Renzini}}]{Daddi2005}
{Daddi}, E., {Dickinson}, M., {Chary}, R., {Pope}, A., {Morrison}, G.,
  {Alexander}, D.~M., {Bauer}, F.~E., {Brandt}, W.~N., {Giavalisco}, M.,
  {Ferguson}, H., {Lee}, K.-S., {Lehmer}, B.~D., {Papovich}, C., \& {Renzini},
  A. 2005, \apjl, 631, L13

\bibitem[{{Dalla Vecchia} \& {Schaye}(2008){Dalla Vecchia} \&
  {Schaye}}]{Dalla2008}
{Dalla Vecchia}, C. \& {Schaye}, J. 2008, \mnras, 387, 1431

\bibitem[{{Dav{\'e}} {et~al.}(2011){Dav{\'e}}, {Finlator}, \&
  {Oppenheimer}}]{Dave2011}
{Dav{\'e}}, R., {Finlator}, K., \& {Oppenheimer}, B.~D. 2011, \mnras, 416, 1354

\bibitem[{{Efstathiou}(2000)}]{Efstathiou2000}
{Efstathiou}, G. 2000, \mnras, 317, 697

\bibitem[{{Farber} {et~al.}(2018){Farber}, {Ruszkowski}, {Yang}, \&
  {Zweibel}}]{Farber2018}
{Farber}, R., {Ruszkowski}, M., {Yang}, H. Y.~K., \& {Zweibel}, E.~G. 2018,
  \apj, 856, 112

\bibitem[{{Ferland} {et~al.}(2013){Ferland}, {Porter}, {van Hoof}, {Williams},
  {Abel}, {Lykins}, {Shaw}, {Henney}, \& {Stancil}}]{Ferland2013}
{Ferland}, G.~J., {Porter}, R.~L., {van Hoof}, P.~A.~M., {Williams}, R.~J.~R.,
  {Abel}, N.~P., {Lykins}, M.~L., {Shaw}, G., {Henney}, W.~J., \& {Stancil},
  P.~C. 2013, \rmxaa, 49, 137

\bibitem[{{F{\"o}rster Schreiber} {et~al.}(2003){F{\"o}rster Schreiber},
  {Genzel}, {Lutz}, \& {Sternberg}}]{Forester2003}
{F{\"o}rster Schreiber}, N.~M., {Genzel}, R., {Lutz}, D., \& {Sternberg}, A.
  2003, \apj, 599, 193

\bibitem[{{Fragile} {et~al.}(2017){Fragile}, {Anninos}, {Croft}, {Lacy}, \&
  {Witry}}]{Fragile2017}
{Fragile}, P.~C., {Anninos}, P., {Croft}, S., {Lacy}, M., \& {Witry}, J.~W.~L.
  2017, \apj, 850, 171

\bibitem[{{Fryxell} {et~al.}(2000){Fryxell}, {Olson}, {Ricker}, {Timmes},
  {Zingale}, {Lamb}, {MacNeice}, {Rosner}, {Truran}, \& {Tufo}}]{Fryxell2000}
{Fryxell}, B., {Olson}, K., {Ricker}, P., {Timmes}, F.~X., {Zingale}, M.,
  {Lamb}, D.~Q., {MacNeice}, P., {Rosner}, R., {Truran}, J.~W., \& {Tufo}, H.
  2000, The Astrophysical Journal Supplement Series, 131, 273

\bibitem[{{Gan} {et~al.}(2018){Gan}, {Choi}, {Ostriker}, {Ciotti}, \&
  {Pellegrini}}]{Gan2018}
{Gan}, Z., {Choi}, E., {Ostriker}, J.~P., {Ciotti}, L., \& {Pellegrini}, S.
  2018, arXiv e-prints

\bibitem[{{Gray} \& {Scannapieco}(2010){Gray} \& {Scannapieco}}]{Gray2010}
{Gray}, W.~J. \& {Scannapieco}, E. 2010, \apj, 718, 417

\bibitem[{{Gray} \& {Scannapieco}(2011{\natexlab{a}}){Gray} \&
  {Scannapieco}}]{Gray2011a}
---. 2011{\natexlab{a}}, \apj, 733, 88

\bibitem[{{Gray} \& {Scannapieco}(2011{\natexlab{b}}){Gray} \&
  {Scannapieco}}]{Gray2011b}
---. 2011{\natexlab{b}}, \apj, 742, 100

\bibitem[{{Gray} \& {Scannapieco}(2016){Gray} \& {Scannapieco}}]{Gray2016}
---. 2016, \apj, 818, 198

\bibitem[{{Gray} \& {Scannapieco}(2017){Gray} \& {Scannapieco}}]{Gray2017}
---. 2017, \apj, 849, 132

\bibitem[{{Gray} {et~al.}(2015){Gray}, {Scannapieco}, \& {Kasen}}]{Gray2015}
{Gray}, W.~J., {Scannapieco}, E., \& {Kasen}, D. 2015, \apj, 801, 107

\bibitem[{{Greggio} {et~al.}(1998){Greggio}, {Tosi}, {Clampin}, {De Marchi},
  {Leitherer}, {Nota}, \& {Sirianni}}]{Greggio1998}
{Greggio}, L., {Tosi}, M., {Clampin}, M., {De Marchi}, G., {Leitherer}, C.,
  {Nota}, A., \& {Sirianni}, M. 1998, \apj, 504, 725

\bibitem[{{Griffiths} {et~al.}(2000){Griffiths}, {Ptak}, {Feigelson},
  {Garmire}, {Townsley}, {Brandt}, {Sambruna}, \& {Bregman}}]{Griffiths2000}
{Griffiths}, R.~E., {Ptak}, A., {Feigelson}, E.~D., {Garmire}, G., {Townsley},
  L., {Brandt}, W.~N., {Sambruna}, R., \& {Bregman}, J.~N. 2000, Science, 290,
  1325

\bibitem[{{Grimes} {et~al.}(2009){Grimes}, {Heckman}, {Aloisi}, {Calzetti},
  {Leitherer}, {Martin}, {Meurer}, {Sembach}, \& {Strickland}}]{Grimes2009}
{Grimes}, J.~P., {Heckman}, T., {Aloisi}, A., {Calzetti}, D., {Leitherer}, C.,
  {Martin}, C.~L., {Meurer}, G., {Sembach}, K., \& {Strickland}, D. 2009,
  \apjs, 181, 272

\bibitem[{{Haardt} \& {Madau}(2012){Haardt} \& {Madau}}]{Haardt2012}
{Haardt}, F. \& {Madau}, P. 2012, \apj, 746, 125

\bibitem[{{Hafen} {et~al.}(2017){Hafen}, {Faucher-Gigu{\`e}re},
  {Angl{\'e}s-Alc{\'a}zar}, {Kere{\v s}}, {Feldmann}, {Chan}, {Quataert},
  {Murray}, \& {Hopkins}}]{Hafen2017}
{Hafen}, Z., {Faucher-Gigu{\`e}re}, C.-A., {Angl{\'e}s-Alc{\'a}zar}, D.,
  {Kere{\v s}}, D., {Feldmann}, R., {Chan}, T.~K., {Quataert}, E., {Murray},
  N., \& {Hopkins}, P.~F. 2017, \mnras, 469, 2292

\bibitem[{{Hayes} {et~al.}(2016){Hayes}, {Melinder}, {{\"O}stlin}, {Scarlata},
  {Lehnert}, \& {Mannerstr{\"o}m-Jansson}}]{Hayes2016}
{Hayes}, M., {Melinder}, J., {{\"O}stlin}, G., {Scarlata}, C., {Lehnert},
  M.~D., \& {Mannerstr{\"o}m-Jansson}, G. 2016, \apj, 828, 49

\bibitem[{{Hayward} \& {Hopkins}(2017){Hayward} \& {Hopkins}}]{Hayward2017}
{Hayward}, C.~C. \& {Hopkins}, P.~F. 2017, \mnras, 465, 1682

\bibitem[{{Heckman} {et~al.}(2017){Heckman}, {Borthakur}, {Wild},
  {Schiminovich}, \& {Bordoloi}}]{Heckman2017}
{Heckman}, T., {Borthakur}, S., {Wild}, V., {Schiminovich}, D., \& {Bordoloi},
  R. 2017, \apj, 846, 151

\bibitem[{{Heckman} {et~al.}(2015){Heckman}, {Alexandroff}, {Borthakur},
  {Overzier}, \& {Leitherer}}]{Heckman2015}
{Heckman}, T.~M., {Alexandroff}, R.~M., {Borthakur}, S., {Overzier}, R., \&
  {Leitherer}, C. 2015, \apj, 809, 147

\bibitem[{{Heckman} {et~al.}(1990){Heckman}, {Armus}, \& {Miley}}]{Heckman1990}
{Heckman}, T.~M., {Armus}, L., \& {Miley}, G.~K. 1990, The Astrophysical
  Journal Supplement Series, 74, 833

\bibitem[{{Heckman} {et~al.}(1995){Heckman}, {Dahlem}, {Lehnert}, {Fabbiano},
  {Gilmore}, \& {Waller}}]{Heckman1995}
{Heckman}, T.~M., {Dahlem}, M., {Lehnert}, M.~D., {Fabbiano}, G., {Gilmore},
  D., \& {Waller}, W.~H. 1995, \apj, 448, 98

\bibitem[{{Heckman} {et~al.}(2000){Heckman}, {Lehnert}, {Strickland}, \&
  {Armus}}]{Heckman2000}
{Heckman}, T.~M., {Lehnert}, M.~D., {Strickland}, D.~K., \& {Armus}, L. 2000,
  The Astrophysical Journal Supplement Series, 129, 493

\bibitem[{{Hopkins} {et~al.}(2012){Hopkins}, {Quataert}, \&
  {Murray}}]{Hopkins2012}
{Hopkins}, P.~F., {Quataert}, E., \& {Murray}, N. 2012, \mnras, 421, 3522

\bibitem[{{Klein} {et~al.}(1994){Klein}, {McKee}, \& {Colella}}]{Klein94}
{Klein}, R.~I., {McKee}, C.~F., \& {Colella}, P. 1994, \apj, 420, 213

\bibitem[{{Lehnert} \& {Heckman}(1996){Lehnert} \& {Heckman}}]{Lehnert96}
{Lehnert}, M.~D. \& {Heckman}, T.~M. 1996, \apj, 462, 651

\bibitem[{{Lehnert} {et~al.}(1999){Lehnert}, {Heckman}, \&
  {Weaver}}]{Lehnert99}
{Lehnert}, M.~D., {Heckman}, T.~M., \& {Weaver}, K.~A. 1999, \apj, 523, 575

\bibitem[{{Leitherer} {et~al.}(1999){Leitherer}, {Schaerer}, {Goldader},
  {Delgado}, {Robert}, {Kune}, {de Mello}, {Devost}, \&
  {Heckman}}]{Leitherer1999}
{Leitherer}, C., {Schaerer}, D., {Goldader}, J.~D., {Delgado}, R. M.~G.,
  {Robert}, C., {Kune}, D.~F., {de Mello}, D.~F., {Devost}, D., \& {Heckman},
  T.~M. 1999, The Astrophysical Journal Supplement Series, 123, 3

\bibitem[{{Lu} {et~al.}(2015){Lu}, {Blanc}, \& {Benson}}]{Lu2015}
{Lu}, Y., {Blanc}, G.~A., \& {Benson}, A. 2015, \apj, 808, 129

\bibitem[{{Mac Low} \& {Ferrara}(1999){Mac Low} \& {Ferrara}}]{Maclow1999}
{Mac Low}, M.-M. \& {Ferrara}, A. 1999, \apj, 513, 142

\bibitem[{{Mac Low} \& {Zahnle}(1994){Mac Low} \& {Zahnle}}]{MacLow1994}
{Mac Low}, M.-M. \& {Zahnle}, K. 1994, \apjl, 434, L33

\bibitem[{{Martin}(1999)}]{Martin1999}
{Martin}, C.~L. 1999, \apj, 513, 156

\bibitem[{{Martin}(2005)}]{Martin2005}
---. 2005, \apj, 621, 227

\bibitem[{{Martin} {et~al.}(2015){Martin}, {Dijkstra}, {Henry}, {Soto},
  {Danforth}, \& {Wong}}]{Martin2015}
{Martin}, C.~L., {Dijkstra}, M., {Henry}, A., {Soto}, K.~T., {Danforth}, C.~W.,
  \& {Wong}, J. 2015, \apj, 803, 6

\bibitem[{{Martin} {et~al.}(2012){Martin}, {Shapley}, {Coil}, {Kornei},
  {Bundy}, {Weiner}, {Noeske}, \& {Schiminovich}}]{Martin2012}
{Martin}, C.~L., {Shapley}, A.~E., {Coil}, A.~L., {Kornei}, K.~A., {Bundy}, K.,
  {Weiner}, B.~J., {Noeske}, K.~G., \& {Schiminovich}, D. 2012, \apj, 760, 127

\bibitem[{{McKeith} {et~al.}(1995){McKeith}, {Greve}, {Downes}, \&
  {Prada}}]{Mckeith1995}
{McKeith}, C.~D., {Greve}, A., {Downes}, D., \& {Prada}, F. 1995, \aap, 293,
  703

\bibitem[{{McQuinn} {et~al.}(2010){McQuinn}, {Skillman}, {Cannon}, {Dalcanton},
  {Dolphin}, {Hidalgo- Rodr{\'\i}guez}, {Holtzman}, {Stark}, {Weisz}, \&
  {Williams}}]{McQuinn2010}
{McQuinn}, K. B.~W., {Skillman}, E.~D., {Cannon}, J.~M., {Dalcanton}, J.,
  {Dolphin}, A., {Hidalgo- Rodr{\'\i}guez}, S., {Holtzman}, J., {Stark}, D.,
  {Weisz}, D., \& {Williams}, B. 2010, \apj, 724, 49

\bibitem[{{Mori} {et~al.}(2002){Mori}, {Ferrara}, \& {Madau}}]{Mori2002}
{Mori}, M., {Ferrara}, A., \& {Madau}, P. 2002, \apj, 571, 40

\bibitem[{{Mukherjee} {et~al.}(2018){Mukherjee}, {Bicknell}, {Wagner},
  {Sutherland}, \& {Silk}}]{Mukherjee2018}
{Mukherjee}, D., {Bicknell}, G.~V., {Wagner}, A.~Y., {Sutherland}, R.~S., \&
  {Silk}, J. 2018, \mnras, 479, 5544

\bibitem[{{Muratov} {et~al.}(2017){Muratov}, {Kere{\v s}},
  {Faucher-Gigu{\`e}re}, {Hopkins}, {Ma}, {Angl{\'e}s-Alc{\'a}zar}, {Chan},
  {Torrey}, {Hafen}, {Quataert}, \& {Murray}}]{Muratov2017}
{Muratov}, A.~L., {Kere{\v s}}, D., {Faucher-Gigu{\`e}re}, C.-A., {Hopkins},
  P.~F., {Ma}, X., {Angl{\'e}s-Alc{\'a}zar}, D., {Chan}, T.~K., {Torrey}, P.,
  {Hafen}, Z.~H., {Quataert}, E., \& {Murray}, N. 2017, \mnras, 468, 4170

\bibitem[{{Muratov} {et~al.}(2015){Muratov}, {Kere{\v{s}}}, {Faucher-
  Gigu{\`e}re}, {Hopkins}, {Quataert}, \& {Murray}}]{Muratov2015}
{Muratov}, A.~L., {Kere{\v{s}}}, D., {Faucher- Gigu{\`e}re}, C.-A., {Hopkins},
  P.~F., {Quataert}, E., \& {Murray}, N. 2015, \mnras, 454, 2691

\bibitem[{{Murray} {et~al.}(2011){Murray}, {M{\'e}nard}, \&
  {Thompson}}]{Murray2011}
{Murray}, N., {M{\'e}nard}, B., \& {Thompson}, T.~A. 2011, \apj, 735, 66

\bibitem[{{Oey} {et~al.}(2017){Oey}, {Herrera}, {Silich}, {Reiter}, {James},
  {Jaskot}, \& {Micheva}}]{Oey2017}
{Oey}, M.~S., {Herrera}, C.~N., {Silich}, S., {Reiter}, M., {James}, B.~L.,
  {Jaskot}, A.~E., \& {Micheva}, G. 2017, \apj, 849, L1

\bibitem[{{Oppenheimer} {et~al.}(2009){Oppenheimer}, {Dav{\'e}}, \&
  {Finlator}}]{Oppenheimer2009}
{Oppenheimer}, B.~D., {Dav{\'e}}, R., \& {Finlator}, K. 2009, \mnras, 396, 729

\bibitem[{{Orlando} {et~al.}(2006){Orlando}, {Bocchino}, {Peres}, {Reale},
  {Plewa}, \& {Rosner}}]{Orlando2006}
{Orlando}, S., {Bocchino}, F., {Peres}, G., {Reale}, F., {Plewa}, T., \&
  {Rosner}, R. 2006, \aap, 457, 545

\bibitem[{{Orlando} {et~al.}(2008){Orlando}, {Bocchino}, {Reale}, {Peres}, \&
  {Pagano}}]{Orlando2008}
{Orlando}, S., {Bocchino}, F., {Reale}, F., {Peres}, G., \& {Pagano}, P. 2008,
  \apj, 678, 274

\bibitem[{{Ott} {et~al.}(2005){Ott}, {Walter}, \& {Brinks}}]{Ott2005}
{Ott}, J., {Walter}, F., \& {Brinks}, E. 2005, \mnras, 358, 1453

\bibitem[{{Pettini} {et~al.}(2002){Pettini}, {Rix}, {Steidel}, {Adelberger},
  {Hunt}, \& {Shapley}}]{Pettini2002}
{Pettini}, M., {Rix}, S.~A., {Steidel}, C.~C., {Adelberger}, K.~L., {Hunt},
  M.~P., \& {Shapley}, A.~E. 2002, \apj, 569, 742

\bibitem[{{Pettini} {et~al.}(2001){Pettini}, {Shapley}, {Steidel}, {Cuby},
  {Dickinson}, {Moorwood}, {Adelberger}, \& {Giavalisco}}]{Pettini2001}
{Pettini}, M., {Shapley}, A.~E., {Steidel}, C.~C., {Cuby}, J.-G., {Dickinson},
  M., {Moorwood}, A. F.~M., {Adelberger}, K.~L., \& {Giavalisco}, M. 2001,
  \apj, 554, 981

\bibitem[{{Piqueras L{\'o}pez} {et~al.}(2012){Piqueras L{\'o}pez}, {Colina},
  {Arribas}, {Alonso- Herrero}, \& {Bedregal}}]{Piqueras2012}
{Piqueras L{\'o}pez}, J., {Colina}, L., {Arribas}, S., {Alonso- Herrero}, A.,
  \& {Bedregal}, A.~G. 2012, \aap, 546, A64

\bibitem[{{Rubin} {et~al.}(2014){Rubin}, {Prochaska}, {Koo}, {Phillips},
  {Martin}, \& {Winstrom}}]{Rubin2014}
{Rubin}, K. H.~R., {Prochaska}, J.~X., {Koo}, D.~C., {Phillips}, A.~C.,
  {Martin}, C.~L., \& {Winstrom}, L.~O. 2014, \apj, 794, 156

\bibitem[{{Rupke} {et~al.}(2005){Rupke}, {Veilleux}, \& {Sanders}}]{Rupke2005}
{Rupke}, D.~S., {Veilleux}, S., \& {Sanders}, D.~B. 2005, The Astrophysical
  Journal Supplement Series, 160, 115

\bibitem[{{Scannapieco}(2017)}]{Scannapieco2017}
{Scannapieco}, E. 2017, \apj, 837, 28

\bibitem[{{Scannapieco} \& {Br{\"u}ggen}(2015){Scannapieco} \&
  {Br{\"u}ggen}}]{Scannapieco2015}
{Scannapieco}, E. \& {Br{\"u}ggen}, M. 2015, \apj, 805, 158

\bibitem[{{Scannapieco} {et~al.}(2002){Scannapieco}, {Ferrara}, \&
  {Madau}}]{Scannapieco2002}
{Scannapieco}, E., {Ferrara}, A., \& {Madau}, P. 2002, \apj, 574, 590

\bibitem[{{Scannapieco} {et~al.}(2001){Scannapieco}, {Thacker}, \&
  {Davis}}]{Scannapieco2001}
{Scannapieco}, E., {Thacker}, R.~J., \& {Davis}, M. 2001, \apj, 557, 605

\bibitem[{{Scannapieco} {et~al.}(2004){Scannapieco}, {Weisheit}, \&
  {Harlow}}]{Scannapieco2004b}
{Scannapieco}, E., {Weisheit}, J., \& {Harlow}, F. 2004, \apj, 615, 29

\bibitem[{{Schneider} \& {Robertson}(2018){Schneider} \&
  {Robertson}}]{Schneider2018A}
{Schneider}, E.~E. \& {Robertson}, B.~E. 2018, \apj, 860, 135

\bibitem[{{Sharp} \& {Bland-Hawthorn}(2010){Sharp} \&
  {Bland-Hawthorn}}]{Sharp2010}
{Sharp}, R.~G. \& {Bland-Hawthorn}, J. 2010, \apj, 711, 818

\bibitem[{{Silich} \& {Tenorio-Tagle}(2017){Silich} \&
  {Tenorio-Tagle}}]{Silich2017}
{Silich}, S. \& {Tenorio-Tagle}, G. 2017, \mnras, 465, 1375

\bibitem[{{Silich} {et~al.}(2003){Silich}, {Tenorio-Tagle}, \&
  {Mu{\~n}oz-Tu{\~n}{\'o}n}}]{Silich2003}
{Silich}, S., {Tenorio-Tagle}, G., \& {Mu{\~n}oz-Tu{\~n}{\'o}n}, C. 2003, \apj,
  590, 791

\bibitem[{{Silich} {et~al.}(2007){Silich}, {Tenorio-Tagle}, \&
  {Mu{\~n}oz-Tu{\~n}{\'o}n}}]{Silich2007}
---. 2007, \apj, 669, 952

\bibitem[{{Silich} {et~al.}(2004){Silich}, {Tenorio-Tagle}, \&
  {Rodr{\'{\i}}guez-Gonz{\'a}lez}}]{Silich2004}
{Silich}, S., {Tenorio-Tagle}, G., \& {Rodr{\'{\i}}guez-Gonz{\'a}lez}, A. 2004,
  \apj, 610, 226

\bibitem[{{Somerville} \& {Primack}(1999){Somerville} \&
  {Primack}}]{Somerville1999}
{Somerville}, R.~S. \& {Primack}, J.~R. 1999, \mnras, 310, 1087

\bibitem[{{Soto} {et~al.}(2012){Soto}, {Martin}, {Prescott}, \&
  {Armus}}]{Soto2012}
{Soto}, K.~T., {Martin}, C.~L., {Prescott}, M.~K.~M., \& {Armus}, L. 2012,
  \apj, 757, 86

\bibitem[{{Spence} {et~al.}(2018){Spence}, {Tadhunter}, {Rose}, \&
  {Rodr{\'{\i}}guez Zaur{\'{\i}}n}}]{Spence2018}
{Spence}, R.~A.~W., {Tadhunter}, C.~N., {Rose}, M., \& {Rodr{\'{\i}}guez
  Zaur{\'{\i}}n}, J. 2018, \mnras, 478, 2438

\bibitem[{{Springel} \& {Hernquist}(2003){Springel} \&
  {Hernquist}}]{Springel2003}
{Springel}, V. \& {Hernquist}, L. 2003, \mnras, 339, 289

\bibitem[{{Strickland} \& {Heckman}(2007){Strickland} \&
  {Heckman}}]{Strickland2007}
{Strickland}, D.~K. \& {Heckman}, T.~M. 2007, \apj, 658, 258

\bibitem[{{Strickland} \& {Heckman}(2009){Strickland} \&
  {Heckman}}]{Strickland2009}
---. 2009, \apj, 697, 2030

\bibitem[{{Strickland} \& {Stevens}(2000){Strickland} \&
  {Stevens}}]{Strickland2000}
{Strickland}, D.~K. \& {Stevens}, I.~R. 2000, \mnras, 314, 511

\bibitem[{{Sturm} {et~al.}(2011){Sturm}, {Gonz{\'a}lez-Alfonso}, {Veilleux},
  {Fischer}, {Graci{\'a}-Carpio}, {Hailey- Dunsheath}, {Contursi}, {Poglitsch},
  {Sternberg}, {Davies}, {Genzel}, {Lutz}, {Tacconi}, {Verma}, {Maiolino}, \&
  {de Jong}}]{Sturm2011}
{Sturm}, E., {Gonz{\'a}lez-Alfonso}, E., {Veilleux}, S., {Fischer}, J.,
  {Graci{\'a}-Carpio}, J., {Hailey- Dunsheath}, S., {Contursi}, A.,
  {Poglitsch}, A., {Sternberg}, A., {Davies}, R., {Genzel}, R., {Lutz}, D.,
  {Tacconi}, L., {Verma}, A., {Maiolino}, R., \& {de Jong}, J.~A. 2011, \apj,
  733, L16

\bibitem[{{Tenorio-Tagle} {et~al.}(2007){Tenorio-Tagle}, {W{\"u}nsch},
  {Silich}, \& {Palous}}]{Tenorio-Tagle2007}
{Tenorio-Tagle}, G., {W{\"u}nsch}, R., {Silich}, S., \& {Palous}, J. 2007,
  \apj, 658, 1196

\bibitem[{{Thompson} {et~al.}(2016){Thompson}, {Quataert}, {Zhang}, \&
  {Weinberg}}]{Thompson2016}
{Thompson}, T.~A., {Quataert}, E., {Zhang}, D., \& {Weinberg}, D.~H. 2016,
  \mnras, 455, 1830

\bibitem[{{Tremonti} {et~al.}(2004){Tremonti}, {Heckman}, {Kauffmann},
  {Brinchmann}, {Charlot}, {White}, {Seibert}, {Peng}, {Schlegel}, {Uomoto},
  {Fukugita}, \& {Brinkmann}}]{Tremonti2004}
{Tremonti}, C.~A., {Heckman}, T.~M., {Kauffmann}, G., {Brinchmann}, J.,
  {Charlot}, S., {White}, S. D.~M., {Seibert}, M., {Peng}, E.~W., {Schlegel},
  D.~J., {Uomoto}, A., {Fukugita}, M., \& {Brinkmann}, J. 2004, \apj, 613, 898

\bibitem[{{Tremonti} {et~al.}(2007){Tremonti}, {Moustakas}, \&
  {Diamond-Stanic}}]{Tremonti2007}
{Tremonti}, C.~A., {Moustakas}, J., \& {Diamond-Stanic}, A.~M. 2007, \apj, 663,
  L77

\bibitem[{{Turk} {et~al.}(2011){Turk}, {Smith}, {Oishi}, {Skory}, {Skillman},
  {Abel}, \& {Norman}}]{Turk2011}
{Turk}, M.~J., {Smith}, B.~D., {Oishi}, J.~S., {Skory}, S., {Skillman}, S.~W.,
  {Abel}, T., \& {Norman}, M.~L. 2011, \apjs, 192, 9

\bibitem[{{Veilleux} {et~al.}(2005){Veilleux}, {Cecil}, \&
  {Bland-Hawthorn}}]{Veilleux2005}
{Veilleux}, S., {Cecil}, G., \& {Bland-Hawthorn}, J. 2005, Annual Review of
  Astronomy and Astrophysics, 43, 769

\bibitem[{{Verner} {et~al.}(1996){Verner}, {Ferland}, {Korista}, \&
  {Yakovlev}}]{Verner1996}
{Verner}, D.~A., {Ferland}, G.~J., {Korista}, K.~T., \& {Yakovlev}, D.~G. 1996,
  \apj, 465, 487

\bibitem[{{Verner} \& {Yakovlev}(1995){Verner} \& {Yakovlev}}]{Verner1995}
{Verner}, D.~A. \& {Yakovlev}, D.~G. 1995, \aaps, 109

\bibitem[{{Walter} {et~al.}(2002){Walter}, {Weiss}, \& {Scoville}}]{Walter2002}
{Walter}, F., {Weiss}, A., \& {Scoville}, N. 2002, \apj, 580, L21

\bibitem[{{Wang}(1995{\natexlab{a}})}]{Wang1995a}
{Wang}, B. 1995{\natexlab{a}}, \apj, 444, 590

\bibitem[{{Wang}(1995{\natexlab{b}})}]{Wang1995b}
---. 1995{\natexlab{b}}, \apjl, 444, L17

\bibitem[{{Weiner} {et~al.}(2009){Weiner}, {Coil}, {Prochaska}, {Newman},
  {Cooper}, {Bundy}, {Conselice}, {Dutton}, {Faber}, {Koo}, {Lotz}, {Rieke}, \&
  {Rubin}}]{Weiner2009}
{Weiner}, B.~J., {Coil}, A.~L., {Prochaska}, J.~X., {Newman}, J.~A., {Cooper},
  M.~C., {Bundy}, K., {Conselice}, C.~J., {Dutton}, A.~A., {Faber}, S.~M.,
  {Koo}, D.~C., {Lotz}, J.~M., {Rieke}, G.~H., \& {Rubin}, K.~H.~R. 2009, \apj,
  692, 187

\bibitem[{{Westmoquette} {et~al.}(2009){Westmoquette}, {Smith}, {Gallagher},
  {Trancho}, {Bastian}, \& {Konstantopoulos}}]{Westmoquette2009}
{Westmoquette}, M.~S., {Smith}, L.~J., {Gallagher}, III, J.~S., {Trancho}, G.,
  {Bastian}, N., \& {Konstantopoulos}, I.~S. 2009, \apj, 696, 192

\bibitem[{{W{\"u}nsch} {et~al.}(2011){W{\"u}nsch}, {Silich}, {Palous},
  {Tenorio-Tagle}, \& {Mu{\~n}oz-Tu{\~n}{\'o}n}}]{Wunsch2011}
{W{\"u}nsch}, R., {Silich}, S., {Palous}, J., {Tenorio-Tagle}, G., \&
  {Mu{\~n}oz-Tu{\~n}{\'o}n}, C. 2011, \apj, 740, 75

\bibitem[{{Xu} {et~al.}(2018){Xu}, {Bian}, {Shen}, {Zuo}, {Fan}, \&
  {Zhu}}]{Xu2018}
{Xu}, F., {Bian}, F., {Shen}, Y., {Zuo}, W., {Fan}, X., \& {Zhu}, Z. 2018,
  \mnras, 480, 345

\bibitem[{{Yukita} {et~al.}(2012){Yukita}, {Swartz}, {Tennant}, {Soria}, \&
  {Irwin}}]{Yukita2012}
{Yukita}, M., {Swartz}, D.~A., {Tennant}, A.~F., {Soria}, R., \& {Irwin}, J.~A.
  2012, \apj, 758, 105

\bibitem[{{Zhang} {et~al.}(2014){Zhang}, {Thompson}, {Murray}, \&
  {Quataert}}]{Zhang2014}
{Zhang}, D., {Thompson}, T.~A., {Murray}, N., \& {Quataert}, E. 2014, \apj,
  784, 93

\bibitem[{{Zhang} {et~al.}(2015){Zhang}, {Thompson}, {Quataert}, \&
  {Murray}}]{Zhang2015}
{Zhang}, D., {Thompson}, T.~A., {Quataert}, E., \& {Murray}, N. 2015, ArXiv
  e-prints

\bibitem[{{Zhang} {et~al.}(2017){Zhang}, {Thompson}, {Quataert}, \&
  {Murray}}]{Zhang2017}
---. 2017, \mnras, 468, 4801

\end{thebibliography}
\end{document}